# Triadic instability of a non-resonant precessing fluid cylinder

Romain Lagrange [a] Patrice Meunier [b] Christophe Eloy [b]

[a] *Department of Mathematics, Massachusetts Institute of Technology, Cambridge, MA 02139, USA*
[b] *Aix-Marseille Université, CNRS, Centrale Marseille, IRPHE UMR 7342, Marseille, France*

**Abstract**

Flows forced by a precessional motion can exhibit instabilities of crucial importance, whether they concern the fuel of a flying object or the liquid core of a telluric planet. So far, stability analyses of these flows have focused on the special case of a resonant forcing. Here, we address the instability of the flow inside a precessing cylinder in the general case. We first show that the base flow forced by the cylinder precession is a superposition of a vertical or horizontal shear flow and an infinite sum of forced modes. We then perform a linear stability analysis of this base flow by considering its triadic resonance with two free Kelvin modes. Finally, we derive the amplitude equations of the free Kelvin modes and obtain an expression of the instability threshold and growth rate.

**Résumé**

**Instabilité triadique d'un fluide dans un cylindre en précession. Les écoulements de précession peuvent exhiber des instabilités dont la compréhension est cruciale, que ce soit pour prédire le mouvement du carburant liquide d'un objet volant, ou le mouvement des noyaux liquides des planètes telluriques. Jusqu'à présent, les analyses de stabilité de ces écoulements se sont focalisées sur le cas particulier d'un forçage à une fréquence de résonance. Ici, nous étudions l'instabilité d'un fluide dans un cylindre en précession, pour une fréquence de forçage quelconque. Premièrement, nous montrons que l'écoulement de base d'un fluide dans un cylindre en précession est une superposition d'un cisaillement vertical ou horizontal et une somme de modes forcés. Ensuite, nous analysons la stabilité de l'écoulement de base en considérant sa résonance triadique avec deux modes de Kelvin libres. Finalement, nous dérivons les équations d'amplitude des modes de Kelvin libres et obtenons une expression du seuil d'instabilité et du taux de croissance.**

*Key words:* Precession ; Kelvin modes ; Triadic Resonance ; Stability

*Mots-clés :* Précession ; Modes de Kelvin ; Résonance triadique ; Stabilité

*Email addresses:* `romain.g.lagrange@gmail.com` (Romain Lagrange), `meunier@irphe.univ-mrs.fr` (Patrice Meunier), `eloy@irphe.univ-mrs.fr` (Christophe Eloy).

## 1. Introduction

Knowing the flow forced by precessional motion is of critical importance in several domains. In aeronautics, liquid propellants contained in flying objects can become resonant for specific geometries of their tank. The resulting flow can then create a destabilizing torque on the objects and dangerously modify their trajectories [1, 2, 3, 4, 5, 6, 7, 8, 9]. Modeling the flow inside a precessing cylinder is thus a necessary step to design tank geometries that avoid these unwanted resonances.

In geophysics, most planets have a motion of slow precession, which is mainly governed by the planet aspect ratio. In the presence of a liquid core, this precessional motion creates a weak forcing that can drastically modify the flow inside the core due to the presence of resonances and critical layers. Flows inside liquid planet cores are of primordial interest to understand the generation of magnetic field by dynamo effect. For the present-day Earth, the magnetic field is likely due to the convection between the hot solid inner core and the colder mantle [10, 11, 12, 13]. However, a magnetic field was present on the early Earth although a solid inner core was not yet present. At that time, other mechanisms have generated and sustained the Earth's magnetic field. Tides (leading to elliptic streamlines) have often thought to be a source of energy sufficient for geodynamo [10, 14], but recently it has also been shown numerically that precession could generate a magnetic field [15, 16] (although this was not clearly proven for the case of the Earth). Besides, there is still some debate as to whether the production of kinetic energy due to precession is sufficient to balance the Ohmic energy loss induced by the magnetic field [17, 18, 19, 20, 21, 22, 23, 24]. In any case, even if precession is not the cause of magnetic field production on Earth, it might be different on other telluric planets.

To study the flow driven by a precessional motion, the cylindrical geometry offers a good alternative to a planet-like spheroidal geometry because of its simplicity. In a precessing cylinder, the base flow is a sum of a shear flow and an infinite set of forced modes [25, 26]. For particular precessional frequencies, a forced mode may become resonant when the height of the cylinder equals an odd number of half wavelengths [27, 28, 29, 30]. In the framework of an inviscid and linear theory, this resonance leads to a divergence of the forced mode amplitude. Viscous effects however may saturate this amplitude to a value scaling as the inverse square root of the Ekman number (due to Ekman layers) [22]. Nonlinear effects can also saturate the amplitude at a value scaling as the cubic root of the forcing [30]. This nonlinear saturation is due to the presence of a strong axisymmetric zonal flow (also called geostrophic flow), which tends to decrease the solid body rotation and thus detune the resonance of the forced mode [31].

When the Ekman number is decreased or the precessing angle is increased above a critical value, a resonant forced mode can become unstable [28, 32, 33, 34, 35, 36]. For small tilt angles, we have shown that this instability is due to a triadic resonance between the resonant forced mode and two free Kelvin modes [37, 38, 39]. However, outside of resonances, the forced modes have a small amplitude and the base flow is not made of a single mode anymore. Here, our principal objective is to perform a stability analysis of the complete base flow (made of a shear flow and a sum of forced modes) in the case of a non-resonant precessing fluid cylinder. We will consider the triadic interaction of the base flow with two free Kelvin modes to determine the conditions of instability and derive an expression for the growth rate.

This paper is organized as follows. Section 2 presents the problem of a precessing cylinder by introducing the governing equations. In this section we determine the off-resonance base flow and discuss the symmetry properties of the problem. In § 3, we develop a linear stability analysis of the complete base flow, based on the mechanism of triadic resonance. We discuss the conditions of resonance, derive the amplitude equations of the instability modes and provide an analytical expression of the instability growth rate and threshold. In § 4 we present an application for a particular case, and determine its stability diagram. Finally, some conclusions are drawn and discussed in the context of the transition to turbulence in precessing flows.

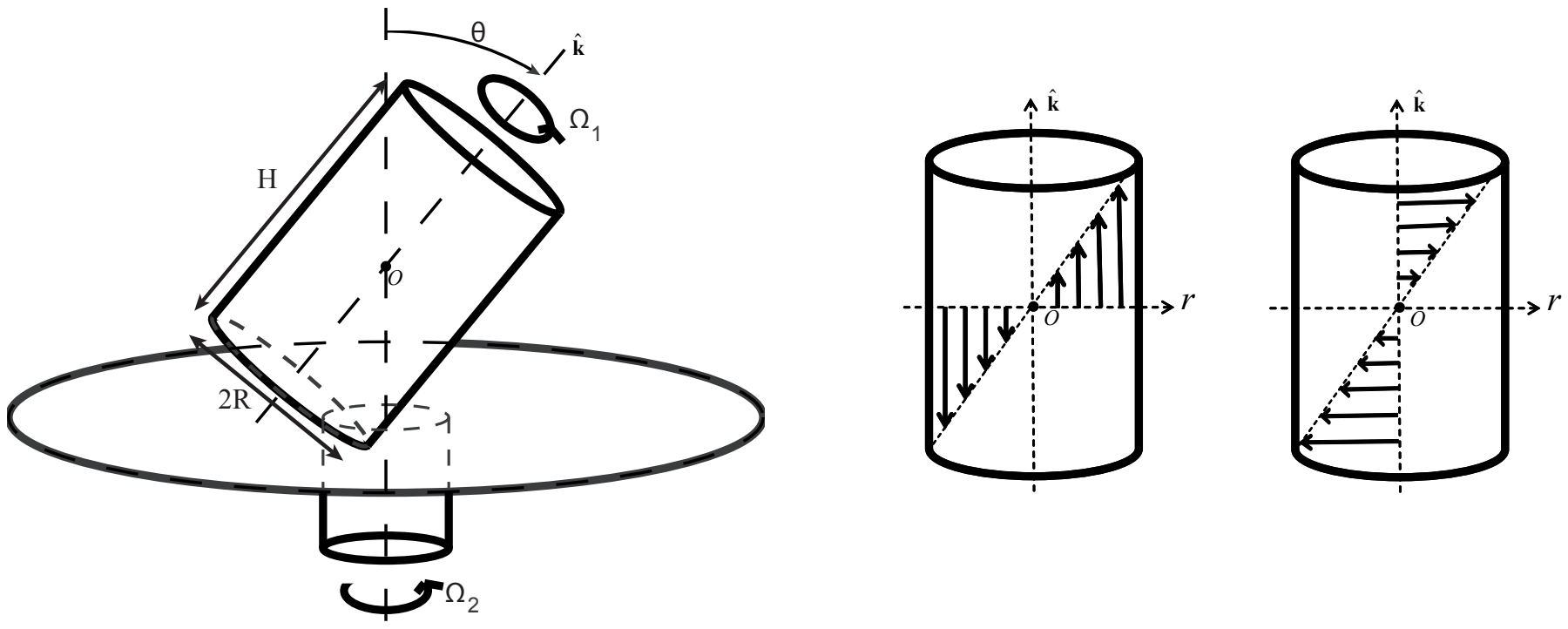

Figure 1. Sketch of a precessing cylinder of radius $R$ and height $H$ (left). The cylinder is rotating around its own axis with angular velocity $\Omega_1$, and this axis itself rotates around a second axis with angular velocity $\Omega_2$. The precession forces a shear flow which can be taken vertical (middle) or horizontal (right), but which does not respect the boundary conditions.

## 2. Formulation of the problem

Consider a cylinder of radius $R$, height $H$, axis of revolution along $\hat{\mathbf{k}}$, entirely filled with a Newtonian fluid of kinematic viscosity $\nu$. The cylinder rotates at the angular speed $\Omega_1$ about $\hat{\mathbf{k}}$, which also rotates at the angular speed $\Omega_2$ about the vertical axis and we denote by $\theta$ the precession angle, i.e. the angle between these two axes of rotation (Fig. 1).

To make the problem dimensionless, we introduce four numbers: the aspect ratio $h = H/R$, the frequency ratio $\omega = \Omega_1/\Omega$, with $\Omega = \Omega_1 + \Omega_2 \cos\theta$, the Ekman number $Ek = \nu/(\Omega R^2)$ and the Rossby number $Ro = \Omega_2 \sin\theta/\Omega$, which will be assumed asymptotically small, i.e. $Ro \ll 1$ (weak precession). The dimensionless flow velocity in the cylinder's frame of reference $(O, \hat{\mathbf{\imath}}, \hat{\mathbf{\jmath}}, \hat{\mathbf{k}})$ is denoted by $\mathbf{u} = \mathbf{U}/(R\Omega)$. The dimensionless cylindrical coordinates are $(r, \varphi, z)$, where $z = 0$ corresponds to the mid-height section of the cylinder and we note $\mathbf{r}$ the position vector of a fluid particle. In the cylinder's frame of reference, the dimensionless Euler equations (assuming an inviscid fluid) are [30, 39]

$$\frac{\partial \mathbf{u}}{\partial t} + 2\left(\hat{\mathbf{k}} + Ro\,\boldsymbol{\delta}\right) \times \mathbf{u} + \boldsymbol{\nabla} p = -2Ro\,\omega r\,\cos(\omega t + \varphi)\,\hat{\mathbf{k}} + \mathbf{u} \times (\boldsymbol{\nabla} \times \mathbf{u}), \tag{1a}$$

$$\boldsymbol{\nabla} \cdot \mathbf{u} = 0, \tag{1b}$$

with $\boldsymbol{\delta} = \cos(\omega t)\,\hat{\mathbf{\imath}} - \sin(\omega t)\,\hat{\mathbf{\jmath}}$. On the left hand side (LHS) of (1a), the first term is inertia, the second term is the Coriolis force and $p$ is the dimensionless pressure field defined as

$$p = \frac{P}{\rho \Omega^2 R^2} - \frac{1}{2} r^2 + Ro|1-\omega| rz \cos(\omega t + \varphi) + \gamma_O \cdot \mathbf{r} - \frac{1}{2} Ro^2 [z^2 + r^2 \sin^2(\omega t + \varphi)] + \frac{1}{2}\mathbf{u}^2, \tag{2}$$

where $\gamma_O = \Gamma_O / R\Omega^2$ is the dimensionless acceleration of the cylinder centroid $O$. On the right hand side (RHS) of (1a), the first term is the forcing due to precession, the second term is the convective nonlinear term. At this point, it is convenient to introduce the four components vector $\mathbf{v} = (\mathbf{u}, p)^T$ and recast equations (1) into a matrix formulation

$$\left(\frac{\partial}{\partial t}\mathcal{I} + \mathcal{M}\right)\mathbf{v} = 2Ro\,\mathbf{F}_0 \cos(\omega t + \varphi) + \mathbf{N}(\mathbf{v}, \mathbf{v}) + Ro\,(\mathcal{D}\ \mathrm{e}^{\mathrm{i}(\omega t + \varphi)} + \text{c.c.})\mathbf{v}, \tag{3}$$

where operators $\mathcal{I}$, $\mathcal{M}$, $\mathcal{D}$, the forcing vector $\mathbf{F}_0$ and the bilinear function $\mathbf{N}$ are reported in Appendix A. The symbol c.c. stands for the complex conjugate.

### 2.1. *Base flow*

In the limit of small Rossby numbers, the base flow forced by the precessional motion can be found by solving the Euler equations (3) at first order. We thus have to solve the following inhomogeneous linear differential equation for $\mathbf{v}$

$$\left(\frac{\partial}{\partial t}\mathcal{I} + \mathcal{M}\right)\mathbf{v} = 2Ro\,\mathbf{F}_0\cos\left(\omega t + \varphi\right). \tag{4}$$

Projecting this equation onto $\hat{\mathbf{k}}$ yields

$$\frac{\partial v_z}{\partial t} + \frac{\partial p}{\partial z} = -2Ro\,r\omega\cos(\omega t + \varphi). \tag{5}$$

There are two particular solutions of this equation, which can be found by assuming either $p = 0$ or $u_z = 0$. The first assumption leads to a vertical shear given by

$$\mathbf{v} = Ro\ \mathbf{v}^{\mathrm{V}}_{\mathrm{shear}}\ e^{\mathrm{i}(\omega t+\varphi)} + \text{c.c.} \qquad \text{with} \qquad \mathbf{v}^{\mathrm{V}}_{\mathrm{shear}} = \begin{pmatrix} 0 \\ 0 \\ \mathrm{i}r \\ 0 \end{pmatrix}. \tag{6}$$

This vertical shear is schematically shown in Fig. 1 (middle) and corresponds to the flow that would be found in a cylinder with infinite height. However, there is an alternative particular solution given by the second assumption ($u_z = 0$) which leads to a horizontal shear flow

$$\mathbf{v} = Ro\ \mathbf{v}^{\mathrm{H}}_{\mathrm{shear}}\ e^{\mathrm{i}(\omega t+\varphi)} + \text{c.c.} \qquad \text{with} \qquad \mathbf{v}^{\mathrm{H}}_{\mathrm{shear}} = \frac{\omega z}{2-\omega}\begin{pmatrix} \mathrm{i} \\ -1 \\ 0 \\ r\,(\omega - 2) \end{pmatrix}. \tag{7}$$

This horizontal shear is schematically shown in Fig. 1 (right). None of these particular solutions satisfy the boundary conditions on the cylinder walls, and they have to be completed with homogeneous solutions of (4). In the case of the vertical shear $\mathbf{v}^{\mathrm{V}}_{\mathrm{shear}}$, the full solution with proper boundary conditions has been found to be (e.g. [30])

$$\mathbf{v} = Ro\,\mathbf{v}^{\mathrm{V}}_{\mathrm{base}} = Ro\left(\mathbf{v}^{\mathrm{V}}_{\mathrm{shear}} + \sum_{j=1}^{\infty} a^{\mathrm{V}}_{j}\mathbf{v}^{\mathrm{V}}_{j}\right)e^{\mathrm{i}(\omega t+\varphi)} + \text{c.c.}, \tag{8}$$

where $\mathbf{v}^{\mathrm{V}}_{j}$ are forced modes of azimuthal wavenumber $m = 1$, frequency $\omega$, and axial wavenumbers $k^{\mathrm{V}}_{j}$. The amplitudes $a^{\mathrm{V}}_{j}$ and the structure of the forced modes are given in Appendix A. Since here, these forced modes compensate the normal flow on the top and bottom walls of the cylinder, they have zero radial velocity $u_r = 0$ at $r = 1$. This boundary condition forces the axial wavenumbers $k^{\mathrm{V}}_{j}$ through the dispersion relation $D(1, \omega, k^{\mathrm{V}}_{j}) = 0$ given in Appendix A (it is stressed that the wavenumbers $k^{\mathrm{V}}_{j}$ are not a multiple of $\pi/h$ since the forced modes do not satisfy the boundary condition $u_z = 0$ on the top and bottom wall).

In the case of a horizontal shear $\mathbf{v}^{\mathrm{H}}_{\mathrm{shear}}$, a similar procedure yields the following base flow solution

$$\mathbf{v} = Ro\,\mathbf{v}^{\mathrm{H}}_{\mathrm{base}} = Ro\left(\mathbf{v}^{\mathrm{H}}_{\mathrm{shear}} + \sum_{j=1}^{\infty} a^{\mathrm{H}}_{j}\mathbf{v}^{\mathrm{H}}_{j}\right)e^{\mathrm{i}(\omega t+\varphi)} + \text{c.c.}. \tag{9}$$

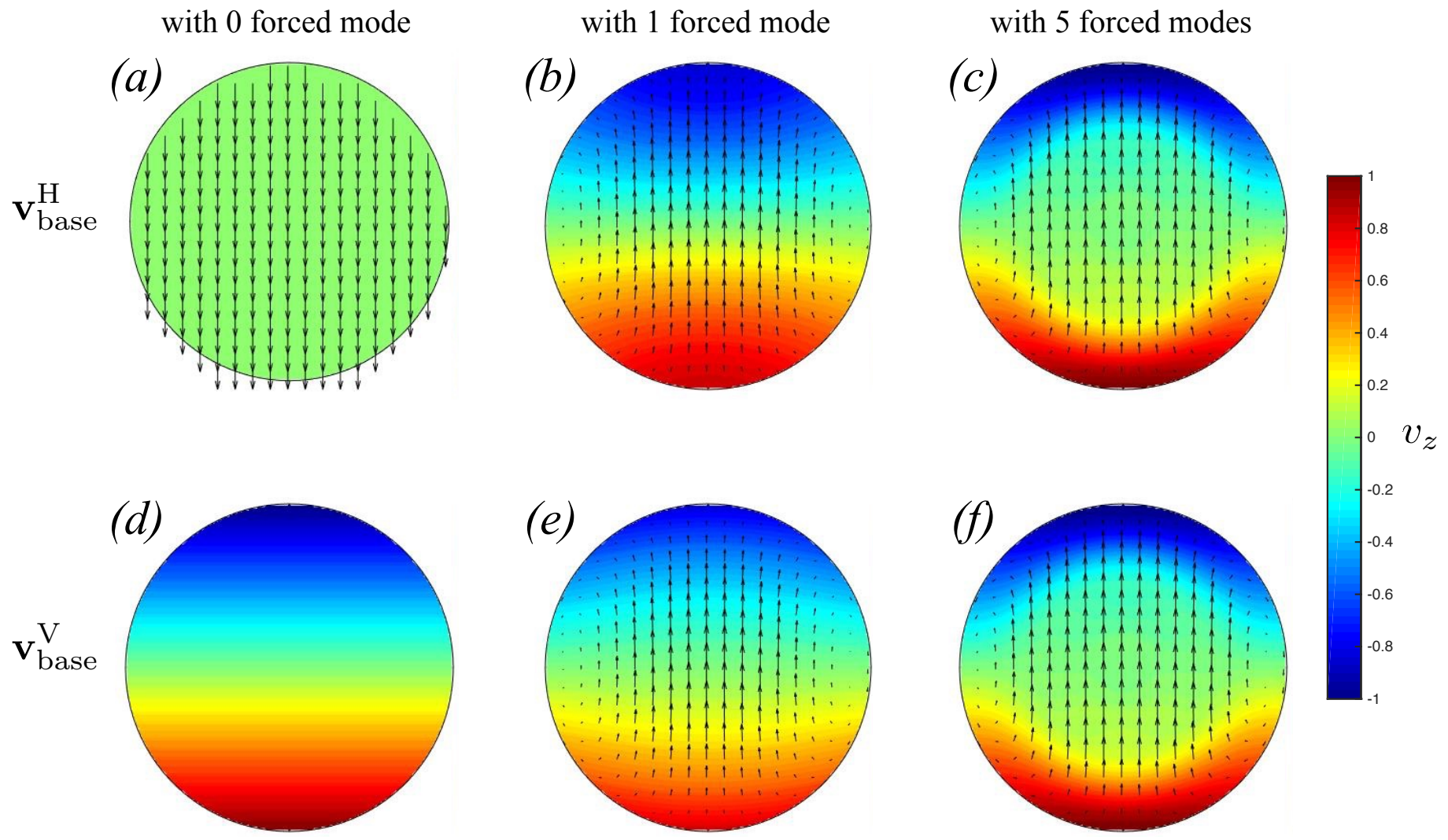

Figure 2. Comparisons of the base flows computed from eqns. (8) and (9) for $h = 2.3$, $\omega = 1.34$, $t = 0$. *(a-c)* Flow $\mathbf{v}^{\mathrm{H}}_{\mathrm{base}}$ obtained from eqn. (9) when the sum of forced modes is truncated to 0, 1, and 5 modes respectively. *(d-f)* Same plots for $\mathbf{v}^{\mathrm{V}}_{\mathrm{base}}$ from eqn. (8). In both cases, the vectors show the projection of the velocity field in the plane $z = h/3$ and the color-coded map shows $v_z$ in the same plane.

The amplitudes $a^{\mathrm{H}}_j$ and the structure of the forced modes $\mathbf{v}^{\mathrm{H}}_j$ are given in Appendix A. In this case, note that the axial wavenumbers $k^{\mathrm{H}}_j$ are odd multiples of $\pi/h$ ($k^{\mathrm{H}}_j = (2j-1)\pi/h$). Here, the forced modes $\mathbf{v}^{\mathrm{H}}_j$ satisfy the boundary conditions on the top and bottom walls, but not on the lateral walls ($u_r \neq 0$ at $r = 1$), because these forced modes have been added to compensate an horizontal shear flow.

In both cases ($\mathbf{v}^{\mathrm{V}}_{\mathrm{base}}$ or $\mathbf{v}^{\mathrm{H}}_{\mathrm{base}}$), the base flow can be written as a superposition of a shear flow (vertical or horizontal respectively) and a sum of forced modes. These forced modes are similar to Kelvin modes [25, 26]. However, we do not consider them as "real" Kelvin modes because they do not satisfy the free-slip boundary condition on all the cylinder walls. Figure 2 shows the two base flow solutions with an increasing number of forced modes. It is clear that both solutions tend to be equal $\mathbf{v}^{\mathrm{V}}_{\mathrm{base}} = \mathbf{v}^{\mathrm{H}}_{\mathrm{base}} = \mathbf{v}_{\mathrm{base}}$ when the number of forced modes is increased.

### 2.2. *Symmetry properties*

The stability analysis of the base flow described above will be presented below in §3. Before that, it is worth recalling some symmetry properties of the problem. It can first be noted that the vertical shear flow $\mathbf{v}^{\mathrm{V}}_{\mathrm{shear}}$ has only a vertical velocity component $v_z$, which is an even function of $z$. In contrast, the horizontal shear flow $\mathbf{v}^{\mathrm{H}}_{\mathrm{shear}}$ has no vertical component but its radial, azimuthal, and pressure components are odd functions of $z$. Both quadrivectors are thus of the type:

$$\mathbf{v}^{-} = \begin{pmatrix} f^{-}(z,r,\varphi,t) \\ f^{-}(z,r,\varphi,t) \\ f^{+}(z,r,\varphi,t) \\ f^{-}(z,r,\varphi,t) \end{pmatrix}, \tag{10}$$

where $f^+$ (resp. $f^-$) denotes even (resp. odd) functions of $z$. Equation (A.3) shows that the forced modes $\mathbf{v}_j^{\mathrm{V}}$ and $\mathbf{v}_j^{\mathrm{H}}$ also have the same $z$-parity. As a consequence, the base flow calculated above is of the type $\mathbf{v}^-$. This could have been anticipated since the first order terms of the Navier–Stokes equations only force this symmetry. However, at higher orders, the operators can generate a flow with the opposite symmetry of the type

$$\mathbf{v}^+ = \begin{pmatrix} f^+(z,r,\varphi,t) \\ f^+(z,r,\varphi,t) \\ f^-(z,r,\varphi,t) \\ f^+(z,r,\varphi,t) \end{pmatrix}. \tag{11}$$

It is easy to show that the operators $\mathcal{D}$, $\overline{\mathcal{D}}$, and $\mathbf{N}$ have the following properties

$$\mathcal{D}\mathbf{v}^- \sim \overline{\mathcal{D}}\mathbf{v}^- \sim \mathbf{v}^+, \tag{12a}$$

$$\mathbf{N}\left(\mathbf{v}_{\text{base}}, \mathbf{v}^-\right) \sim \mathbf{N}\left(\overline{\mathbf{v}_{\text{base}}}, \mathbf{v}^-\right) \sim \mathbf{v}^+, \tag{12b}$$

$$\mathbf{N}\left(\mathbf{v}^-, \mathbf{v}_{\text{base}}\right) \sim \mathbf{N}\left(\mathbf{v}^-, \overline{\mathbf{v}_{\text{base}}}\right) \sim \mathbf{v}^+, \tag{12c}$$

where the symbol $\sim$ means "same parity as" and $\mathbf{v}_{\text{base}}$ corresponds to either the vertical base flow $\mathbf{v}_{\text{base}}^{\mathrm{V}}$ or the horizontal base flow $\mathbf{v}_{\text{base}}^{\mathrm{H}}$. This means that, even if the base flow is of the type $\mathbf{v}^-$ at first order, the nonlinear terms may introduce a different symmetry in the problem. It is actually easy to show that the previous equations remain valid under the permutation of signs $(+,-) \to (-,+)$.

A direct consequence of these symmetry properties is that a triadic resonance cannot develop if the perturbation has a single parity ($\mathbf{v}^+$ or $\mathbf{v}^-$). Indeed, in a mechanism of triadic resonance, the base flow interacts with two free Kelvin modes $\mathbf{v}_1$ and $\mathbf{v}_2$ through the nonlinear operator $\mathbf{N}$. Let us assume that the free Kelvin modes have the same symmetry $\mathbf{v}^-$. The growth of the first free Kelvin mode is due to the nonlinear interaction of the second free Kelvin mode with the base flow, via the terms $\mathbf{N}(\mathbf{v}_{\text{base}}, \mathbf{v}_2^-)$, $\mathbf{N}(\mathbf{v}_2^-, \mathbf{v}_{\text{base}})$ and $\mathcal{D}\mathbf{v}_2^-$ which have the opposite symmetry $\mathbf{v}^+$. These forcing terms are thus perpendicular to the first free Kelvin mode and such a triadic resonance is non constructive. This can be properly shown by defining the dot product

$$\langle \mathbf{X}, \mathbf{Y} \rangle = \int_V \left(\overline{X_r}Y_r + \overline{X_\varphi}Y_\varphi + \overline{X_z}Y_z + \overline{X_p}Y_p\right) d^3V, \tag{13}$$

where $\overline{X}$ refers to the conjugate of $X$ and $V$ is the volume of the cylinder. It is then straightforward to show, using (12a-c) that the dot products $\langle \mathbf{v}_1^-, \mathbf{N}(\mathbf{v}_2^-, \mathbf{v}_{\text{base}})\rangle$, $\langle \mathbf{v}_1^-, \mathbf{N}(\mathbf{v}_{\text{base}}, \mathbf{v}_2^-)\rangle$ and $\langle \mathbf{v}_1^-, \mathcal{D}\mathbf{v}_2^-\rangle$ vanish because they only contain terms of the form

$$\langle \mathbf{v}^+, \mathbf{v}^- \rangle = \int_{-h/2}^{h/2} f^+(z) f^-(z) dz = 0. \tag{14}$$

The same reasoning applies to free Kelvin modes with a symmetry $\mathbf{v}^+$. The general conclusion is that the constructive triadic resonances must couple an even free Kelvin mode $\mathbf{v}^+$ with an odd free Kelvin mode $\mathbf{v}^-$. We will now use this property to restrict the number of possible instabilities that may arise in the linear stability analysis.

## 3. Linear stability analysis

To study the stability of the base flow, we introduce a small perturbation in form of a four-components vector $\widetilde{\mathbf{v}} = (\widetilde{\mathbf{u}}, \widetilde{p})^T$, so that the total flow is

$$\mathbf{v} = Ro\,\mathbf{v}_{\text{base}} + \widetilde{\mathbf{v}} + o\,(Ro)\,, \tag{15}$$

where $\mathbf{v}_{\text{base}}$ is either $\mathbf{v}^{\text{V}}_{\text{base}}$ or $\mathbf{v}^{\text{H}}_{\text{base}}$. Inserting this expansion into (3) yields an equation for the perturbation vector

$$\left(\frac{\partial}{\partial t}\mathcal{I} + \mathcal{M}\right)\widetilde{\mathbf{v}} = Ro\left[\mathbf{N}(\mathbf{v}_{\text{base}}, \widetilde{\mathbf{v}}) + \mathbf{N}(\widetilde{\mathbf{v}}, \mathbf{v}_{\text{base}}) + \left(\mathcal{D}\ \mathrm{e}^{\mathrm{i}(\omega t+\varphi)} + \text{c.c.}\right)\widetilde{\mathbf{v}}\right] + o\,(Ro) + o\,(|\widetilde{\mathbf{v}}|)\,, \tag{16}$$

where $|\widetilde{\mathbf{v}}| = \sqrt{\langle \widetilde{\mathbf{v}}, \widetilde{\mathbf{v}} \rangle}$ is the magnitude of $\widetilde{\mathbf{v}}$. The first two terms on the RHS of (16) represent the nonlinear interactions between the base flow and the perturbation. The third term represents the interaction between the forcing due to precession and the perturbation. The perturbation vector satisfies the inviscid boundary condition

$$\widetilde{\mathbf{u}} \cdot \mathbf{n} = 0 \text{ at the walls } (r = 1 \text{ or } z = \pm h/2)\,. \tag{17}$$

To solve (16) and (17), we use a multiscale expansion in time with $t$ a rapid time scale and $\tau = Ro\, t$ a slow time scale. We then expand $\widetilde{\mathbf{v}}$ as

$$\widetilde{\mathbf{v}} = \widetilde{\mathbf{v}}_0\,(\mathbf{r}, \tau, t) + Ro\,\widetilde{\mathbf{v}}_1\,(\mathbf{r}, \tau, t) + o\,(Ro)\,. \tag{18}$$

Inserting (18) into (16) yields two equations: one of order 1 and one of order $Ro$ that shall be studied now. The equation at order one gives the form of the free Kelvin modes, and the equation at order $Ro$ gives their slow time dynamics, hence their stability properties.

### 3.1. *Order 1: free Kelvin modes*

At first order, the equation (16) and the inviscid boundary condition $\widetilde{\mathbf{u}} \cdot \mathbf{n}$ are

$$\left(\frac{\partial \mathcal{I}}{\partial t} + \mathcal{M}\right)\widetilde{\mathbf{v}}_0 = \mathbf{0}, \tag{19a}$$

$$\widetilde{\mathbf{u}}_0 \cdot \mathbf{n} = 0 \quad \text{at the walls } (r = 1 \text{ or } z = \pm h/2). \tag{19b}$$

The solution to this homogenous problem is a linear combination of free Kelvin modes with different $z$-parities [39]

$$\widetilde{\mathbf{v}}_0 = \sum_{l=1}^{\infty} A_l^{+}\mathbf{v}_l^{+} e^{\mathrm{i}(\omega_l t + m_l \varphi)} + \sum_{l=1}^{\infty} A_l^{-}\mathbf{v}_l^{-} e^{\mathrm{i}(\omega_l t + m_l \varphi)} + \text{c.c.} \tag{20}$$

Vectors $\mathbf{v}_l^{+}$ (resp. $\mathbf{v}_l^{-}$) have axial wavenumbers $k_l^{+}$ (resp. $k_l^{-}$) which are even (resp. odd) multiple of $\pi/h$ in order to respect the condition of no normal flow at the top and bottom ($z = \pm h/2$). This property is interesting because the wavenumbers are separated into two families, which will restrict the number of possible triadic resonances. The components of the free Kelvin modes are given in Appendix A. In (20), $A_l^{\pm}$, $m_l$, and $\omega_l$ are the amplitude, azimuthal wavenumber, and angular frequency of the free Kelvin mode $\mathbf{v}_l^{\pm} e^{\mathrm{i}(\omega_l t + m_l \varphi)}$. The wavenumbers are connected through the dispersion relation $D(m_l, \omega_l, k_l^{\pm})$ such that the radial velocity of the mode vanishes at the cylinder wall $r = 1$.

To examine the mechanism of triadic resonance, the perturbation $\widetilde{\mathbf{v}}_0$ is reduced to a combination of two free Kelvin modes $\mathbf{v}_1$ and $\mathbf{v}_2$ with unknown amplitudes $A_1\,(\tau)$ and $A_2\,(\tau)$

$$\widetilde{\mathbf{v}}_0 = A_1 \mathbf{v}_1 e^{\mathrm{i}(\omega_1 t + m_1 \varphi)} + A_2 \mathbf{v}_2 e^{\mathrm{i}(\omega_2 t + m_2 \varphi)} + \text{c.c.}. \tag{21}$$

From now on, we attribute index 2 to the mode with the highest azimuthal wavenumber (i.e. $m_2 > m_1$).

### 3.2. *Triadic resonance*

We know from the symmetry properties presented in §2.2 that a triadic resonance between the base flow and the two free Kelvin modes can be constructive only if it involves free Kelvin modes with different $z$-parities. Therefore, the wavenumbers $k_1$ and $k_2$ must be integer multiple of $\pi/h$ with different parities. It follows that the difference between the two wavenumbers must be an odd multiple of $\pi/h$:

$$k_2 - k_1 = (2p-1)\pi/h, \tag{22}$$

with $p$ an integer.

In addition, the base flow will interact with two free Kelvin modes if the operator $\mathbf{N}$ appropriately couples their time and azimuthal Fourier components. Thus the coupling term $\mathbf{N}(\mathbf{v}_{\text{base}}, \widetilde{\mathbf{v}}_0) + \mathbf{N}(\widetilde{\mathbf{v}}_0, \mathbf{v}_{\text{base}})$ in (16) has to contain the same Fourier components as $\mathbf{v}_l e^{\mathrm{i}(\omega_l t + m_l \varphi)}$, $l = 1, 2$. Since these terms have the following time and azimuthal Fourier components

$$\mathbf{v}_l e^{\mathrm{i}(\omega_l t + m_l \varphi)} : \text{Fourier components } (m_l, \omega_l), \tag{23a}$$

$$\mathbf{N}(\mathbf{v}_{\text{base}}, \widetilde{\mathbf{v}}_0) + \mathbf{N}(\widetilde{\mathbf{v}}_0, \mathbf{v}_{\text{base}}) : \text{Fourier components } (m_l + 1, \omega_l + \omega), (m_l - 1, \omega_l - \omega), \tag{23b}$$

the base flow will interact with the two free Kelvin modes if

$$m_2 - m_1 = 1, \tag{24a}$$

$$\omega_2 - \omega_1 = \omega. \tag{24b}$$

We recognize on the RHS the azimuthal wavenumber $m_{\text{base}} = 1$ and the angular frequency $\omega_{\text{base}} = \omega$ of the base flow. The conditions (24a-b) are characteristic of triadic resonances occurring in various domains (surface waves, plate vibrations, etc.), and are the key ingredient of weak (or wave) turbulence theory.

To find a pair of free Kelvin modes that fulfill the conditions of resonance (24a-b), we proceed as shown in Fig 3. In the plane $(k_2, \omega_2)$ we plot the dispersion relation of the free Kelvin modes $(m_2, \omega_2, k_2)$ and the dispersion relation of the free Kelvin modes $(m_1, \omega_1, k_1)$ translated horizontally by $(2p-1)\pi/h$ (with $p$ an arbitrary integer) and vertically by $\omega$. The intersection points correspond to free Kelvin modes satisfying the conditions of resonance (24a-b) and the condition induced by parity (22). In addition, the free Kelvin modes have to fit in the cylinder, so their axial wavenumbers need to be multiple of $\pi/h$. It means that the intersection point has to lie on the vertical dotted line of Fig 3. Such a tuned triadic resonance only occurs for particular aspect ratios (for a given forcing frequency $\omega$). Otherwise, in most cases, the three curves do not intersect at the same point. Figure 3 shows an example of a tuned triadic resonance for $h = 2.3$, $\omega = 1.34$ with $m_1 = 2$ and $m_2 = 3$. The label for these points is $(m_2, l_1, l_2)$, where $l_{1,2}$ is the branch number of the dispersion relations. There are an infinity of possible triadic resonances and they can all be studied in the inviscid case. However, when viscous effects are taken into account, the highest wavenumbers will be damped, such that in practice only the lowest axial, azimuthal and radial wavenumbers may be treated.

The procedure outline above shows how the axial, azimuthal and radial wavenumbers of two free Kelvin modes can be found to permit triadic resonance. In the next section, we consider the Euler equations at the next order in *Ro*, in order to calculate the slow temporal evolution of these free Kelvin modes and determine their stability.

### 3.3. *Order Ro: slow time equations*

At order *Ro*, equation (16) becomes

$$\left(\frac{\partial \mathcal{I}}{\partial t} + \mathcal{M}\right)\widetilde{\mathbf{v}}_1 = \mathbf{N}(\mathbf{v}_{\text{base}}, \widetilde{\mathbf{v}}_0) + \mathbf{N}(\widetilde{\mathbf{v}}_0, \mathbf{v}_{\text{base}}) + \left[\left(\mathcal{D}e^{\mathrm{i}(\omega t + \varphi)} + \text{c.c.}\right) - \frac{\partial \mathcal{I}}{\partial \tau}\right]\widetilde{\mathbf{v}}_0. \tag{25}$$

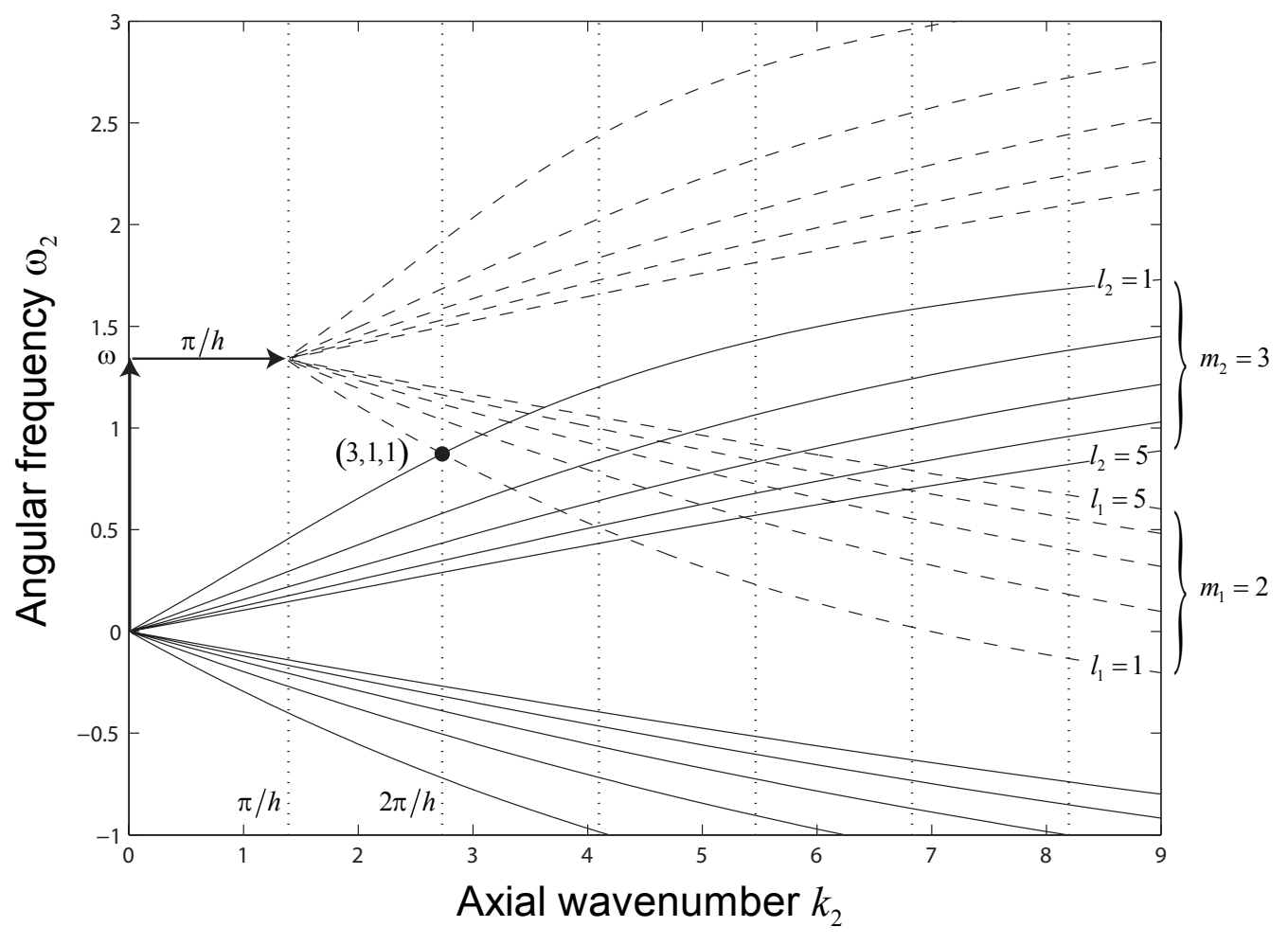

Figure 3. Dispersion relations of the free Kelvin modes with azimuthal wavenumbers $m_1 = 2$ (dashed lines) and $m_2 = 3$ (solid lines). The dispersion relation for $m_1 = 2$ is translated by $\pi/h$ ($h = 2.3$) along the abscissae and by $\omega = 1.34$ along the ordinate. Vertical dotted lines indicate the discretisation of the axial wavenumber as a multiple of $\pi/h$ imposed by the inviscid boundary condition at the top and bottom of the cylinder. A point lying at the intersection of the branches of the dispersion relations and a vertical line corresponds to a pair of resonant free Kelvin modes. The combination $(m_2 = 3, l_1 = 1, l_2 = 1)$ (marked with a black circle) is an example of resonant free Kelvin modes. Coordinates for this point are $k_2 = 2\pi/h$ and $\omega_2 = 0.874$.

This $O(Ro)$ problem is linear, with a forcing term given by the RHS of (25). To avoid secular terms in the solution $\widetilde{\mathbf{v}}_1$, the RHS must be orthogonal to the kernel of the LHS operator. This kernel being spanned by the free Kelvin modes, themselves given by the $O(1)$ problem solved above, a solvability condition is obtained by taking the dot product of (25) with $\mathbf{v}_l e^{\mathrm{i}(\omega_l t + m_l \varphi)}$, $l = 1, 2$. Since the problem is self-adjoint, i.e. $\left\langle \mathbf{v}_l e^{\mathrm{i}(\omega_l t + m_l \varphi)}, (\partial \mathcal{I}/\partial t + \mathcal{M})\, \widetilde{\mathbf{v}}_1 \right\rangle = 0$, we show in Appendix B that the slow time equations for $A_1$ and $A_2$ are

$$\frac{dA_1}{d\tau} = c_1 A_2 \qquad \text{with} \qquad c_1 = \frac{\overline{d}_{12} + n_{1s} + \sum_j \overline{a_j} n_{1j}}{\langle \mathbf{v}_1, \mathcal{I}\mathbf{v}_1 \rangle}, \tag{26a}$$

$$\frac{dA_2}{d\tau} = c_2 A_1 \qquad \text{with} \qquad c_2 = \frac{d_{21} + n_{2s} + \sum_j a_j n_{2j}}{\langle \mathbf{v}_2, \mathcal{I}\mathbf{v}_2 \rangle}. \tag{26b}$$

The terms $dA_l/d\tau$ come from the dot products $\left\langle \mathbf{v}_l e^{\mathrm{i}(\omega_l t + m_l \varphi)}, \frac{\partial \mathcal{I}}{\partial \tau} \widetilde{\mathbf{v}}_0 \right\rangle$. The terms $\overline{d}_{12}$ and $d_{21}$ represent the interaction between the resonant free Kelvin modes and the forcing due to precession. They come from the dot products

$$\left\langle \mathbf{v}_l e^{\mathrm{i}(\omega_l t + m_l \varphi)}, \left( \mathcal{D} e^{\mathrm{i}(\omega t + \varphi)} + \text{c.c.} \right) \widetilde{\mathbf{v}}_0 \right\rangle. \tag{27}$$

The terms $n_{1s}$ and $n_{2s}$ represent the nonlinear interactions between the resonant free Kelvin modes and the shear part of the base flow. They come from the dot products

$$n_{ls} = \left\langle \mathbf{v}_l e^{\mathrm{i}(\omega_l t + m_l \varphi)}, \mathbf{N}\left( \mathbf{v}_s e^{\mathrm{i}(\omega t + \varphi)} + \text{c.c.}, \widetilde{\mathbf{v}}_0 \right) + \mathbf{N}\left( \widetilde{\mathbf{v}}_0, \mathbf{v}_s e^{\mathrm{i}(\omega t + \varphi)} + \text{c.c.} \right) \right\rangle. \tag{28}$$

where $\mathbf{v}_s$ may be the vertical shear given by (6) or the horizontal shear given by (7). Finally, the terms $n_{1j}$ and $n_{2j}$ represent the nonlinear interactions between the resonant free Kelvin modes and the $j$-th forced mode of the base flow. They come from the dot products

$$n_{lj} = \left\langle \mathbf{v}_l e^{\mathrm{i}(\omega_l t + m_l \varphi)}, \mathbf{N}\left( \mathbf{v}_j e^{\mathrm{i}(\omega t + \varphi)} + \text{c.c.}, \widetilde{\mathbf{v}}_0 \right) + \mathbf{N}\left( \widetilde{\mathbf{v}}_0, \mathbf{v}_j e^{\mathrm{i}(\omega t + \varphi)} + \text{c.c.} \right) \right\rangle. \tag{29}$$

Simplified expressions for all these terms are given in Appendix B.

Seeking solutions to the amplitude equations (26) as growing exponentials, $A_j \sim e^{\sigma t} \sim e^{\sigma\tau/Ro}$, yields an analytical prediction for the complex growth rate $\sigma$ of the instability

$$\sigma = |Ro|\sqrt{c_1 c_2}. \tag{30}$$

### 3.4. *Amplitude equations using the horizontal shear decomposition*

Amplitude equations (26) apply to any decomposition, $\mathbf{v}^{\mathrm{V}}_{\mathrm{base}}$ or $\mathbf{v}^{\mathrm{H}}_{\mathrm{base}}$, of the base flow. If the vertical shear decomposition $\mathbf{v}^{\mathrm{V}}_{\mathrm{base}}$ is chosen for the base flow, the formula for the growth rate contains infinite summations (corresponding to $n_{1j}$ and $n_{2j}$). This is because, in this case, axial wavenumbers $k_j^{\mathrm{V}}$ are not integer multiples of $\pi/h$ and the integral from $z=-\pi/h$ to $\pi/h$ in the dot products (29) cannot be easily simplified.

However, if the horizontal decomposition $\mathbf{v}^{\mathrm{H}}_{\mathrm{base}}$ is chosen, it is possible to achieve such a simplification. In this case, the forced modes have axial wavenumbers $k_j^{\mathrm{H}}$ that are odd multiples of $\pi/h$. These forced modes will interact with the resonant free Kelvin modes if the dot product between $\mathbf{N}\left(\mathbf{v}_j^{\mathrm{H}} e^{\mathrm{i}(\omega t+\varphi)} + \text{c.c.}, \widetilde{\mathbf{v}}_0\right)$ and $\mathbf{v}_l e^{\mathrm{i}(\omega_l t + m_l \varphi)}$ lead to a non-zero integral over $z$. Since these terms have the following $z$-Fourier components

$$\mathbf{v}_l : z\text{-Fourier components } \pm k_l, \tag{31a}$$

$$\mathbf{N}\left(\mathbf{v}_j^{\mathrm{H}} e^{\mathrm{i}(\omega t+\varphi)} + \text{c.c.}, \widetilde{\mathbf{v}}_0\right) : z\text{-Fourier components } k_j^{\mathrm{H}} \pm k_l, -k_j^{\mathrm{H}} \pm k_l, \tag{31b}$$

and since $k_1$ and $k_2$ have different parities and $k_j^{\mathrm{H}}$ is an odd multiple of $\pi/h$, the axial wavenumber of the dot product only contains even multiple of $\pi/h$, i.e. multiple of $2\pi/h$. As a consequence, the integral over $z$ is non-zero only if the axial wavenumber of the dot product is equal to zero. It follows that only two forced modes of $\mathbf{v}^{\mathrm{H}}_{\mathrm{base}}$ will interact with the resonant free Kelvin modes and their axial wavenumbers are

$$k_j^{\mathrm{H}} = k_{j_1} = |k_2 - k_1| \quad \text{and} \quad k_j^{\mathrm{H}} = k_{j_2} = |k_2 + k_1|. \tag{32}$$

It means that there are only two forced modes $j_1$ and $j_2$ which give non zero coefficients $n_{1j}$ and $n_{2j}$ such that the infinite summations in (26) becomes sum of only two terms. The coefficients $c_1$ and $c_2$ thus simplify to

$$c_1 = \frac{\overline{d}_{12} + n_{1s} + \overline{a^{\mathrm{H}}_{j_1}} n_{1j_1} + \overline{a^{\mathrm{H}}_{j_2}} n_{1j_2}}{\langle \mathbf{v}_1, \mathcal{I}\mathbf{v}_1\rangle}, \tag{33a}$$

$$c_2 = \frac{d_{21} + n_{2s} + a^{\mathrm{H}}_{j_1} n_{2j_1} + a^{\mathrm{H}}_{j_2} n_{2j_2}}{\langle \mathbf{v}_2, \mathcal{I}\mathbf{v}_2\rangle}. \tag{33b}$$

### 3.5. *Case of a resonant cylinder*

In a resonant cylinder, the precession forces a flow that is dominated by a single forced mode, which is a Kelvin mode (i.e. a forced mode which satisfies the no-slip boundary conditions on the cylinder walls). In that case, the flow can become unstable if this forced Kelvin mode interacts constructively with two free Kelvin modes. As we have shown in a previous work, amplitude equations can be derived for a resonant cylinder [39]. In this section, we want to explain how to recover these amplitude equations from the amplitude equations of a non-resonant precessing fluid cylinder derived here in (26a,b) .

When a forced mode is resonant, its amplitude $a_j$ predicted by the linear theory diverges. As shown by the equation (A.6), it happens when the dispersion relation $D(1,\omega,k_j)=0$ holds for an axial wavenumber $k_j$ which is an odd multiple of $\pi/h$. For large Ekman numbers, the viscous effects saturate the amplitude of the forced Kelvin mode to an order $Ek^{-1/2}$ larger than the amplitudes of the shear flow and the others forced modes, see refs.[22, 30]. Thus, at main order the base flow $\mathbf{v}_{\text{base}}$ is a single forced Kelvin mode with amplitude $|\varepsilon| = O(Ro\,Ek^{-1/2})$. It follows that the summations in (26) are truncated to the index of that mode and the amplitude $a_j$ must be replaced by $\varepsilon/Ro$. Since the terms $\overline{d}_{12}$, $n_{1s}$, $d_{21}$ and $n_{2s}$ are of order $O(1)$, they are negligible compared to $\varepsilon/Ro$ and can be dropped in (26). The amplitude equations when the $j$-th forced mode is resonant thus become

$$\frac{dA_1}{d\tau} = A_2 \frac{(\overline{\varepsilon}/Ro)\, n_{1j}}{\langle \mathbf{v}_1, \mathcal{I}\mathbf{v}_1 \rangle} = n_1\,(\overline{\varepsilon}/Ro)\, A_2, \tag{34a}$$

$$\frac{dA_2}{d\tau} = A_1 \frac{(\varepsilon/Ro)\, n_{2j}}{\langle \mathbf{v}_2, \mathcal{I}\mathbf{v}_2 \rangle} = n_2\,(\varepsilon/Ro)\, A_1. \tag{34b}$$

As explained in §3.4, terms $n_1$ and $n_2$ are non-zeros only if the condition of resonance $|k_2 - k_1| = k_j = (2j-1)\pi/h$ or $|k_2 + k_1| = k_j = (2j-1)\pi/h$ is satisfied (together with similar conditions on $m$ and $\omega$: $m_2 - m_1 = 1$ and $\omega_2 - \omega_1 = \omega$).

Seeking solutions to the amplitude equations (34a,b) as growing exponentials $A_j \sim e^{\sigma t} \sim e^{\sigma\tau/Ro}$, yields an expression for the complex growth rate

$$\sigma = |\varepsilon|\sqrt{n_1 n_2}, \tag{35}$$

similar to the one obtained in [39]. As expected (based on similarities with the elliptic instability), the growth rate scales as the amplitude of the forced Kelvin mode. For large Ekman number, the growth rate of a resonant precessing fluid cylinder is thus an order $Ek^{-1/2}$ larger than the growth rate of the non-resonant case.

For example, for $h = 1.62$ and $\omega = 1.18$, the first Kelvin mode is forced at its first resonance and we recover (see Table 3 in [39]) that the resonant combination $(6, 1, 1)$ has $n_1 = -1.672$, $n_2 = -2.456$, leading to a growth rate $\sigma = 2.026|\varepsilon|$.

### 3.6. *Including viscous and detuning effects*

Amplitude equations (26a,b) have been derived under the assumption of an inviscid fluid and an exactly resonant combination. Accounting for viscosity and free Kelvin modes that do not exactly satisfy the boundary conditions, the amplitude equations modify to

$$\frac{dA_1}{d\tau} = c_1 A_2 - (\alpha_1/Ro)\, A_1, \tag{36a}$$

$$\frac{dA_2}{d\tau} = c_2 A_1 - (\alpha_2/Ro)\, A_2, \tag{36b}$$

with $\alpha_l = s_l Ek^{1/2} + v_l Ek + \mathrm{i} q_l \Delta k_l$ and where $\Delta k_l$ represents the relative distance between a crossing point in Fig. 3 and the closest vertical line [see 39, for more details]. The coefficients $s_l$ represent the surface viscous damping of the free Kelvin modes due to Ekman layers on the cylinder walls. They are complex numbers with a positive real part. The coefficients $v_l$ are real and represent the volume viscous damping of the free Kelvin modes. They come from the Laplace operator of the Navier-Stokes equations, and are proportional to $k_l^2 + \delta_l^2$, such that they strongly damp the free Kelvin modes with large axial and radial wavenumbers . The coefficients $q_l$ are real and represent the damping of the two free Kelvin modes due to detuning effects. All these coefficients are given in [39]. Finally, when the free Kelvin modes are not exactly resonant, the equations (33) do not hold anymore and one has to use the equations (26) to compute the instability terms $c_l$.

Assuming that the amplitudes are growing exponentials, i.e. $A_j \sim e^{\sigma t} \sim e^{\sigma\tau/Ro}$, the equation for the complex growth rate $\sigma$ is obtained by canceling the determinant of (36), leading to

$$(\sigma + \alpha_1)(\sigma + \alpha_2) = Ro^2 c_1 c_2. \tag{37}$$

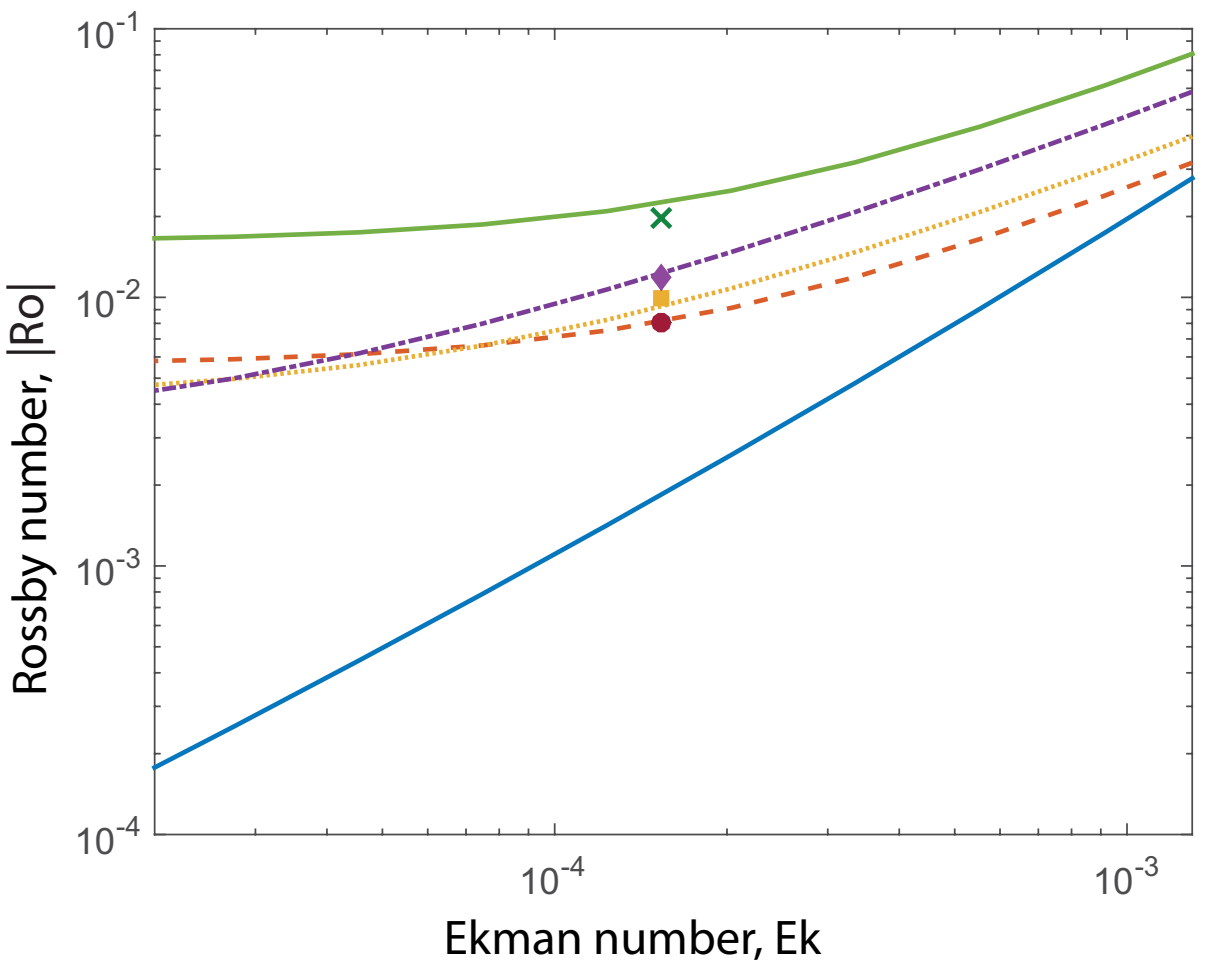

Figure 4. Stability diagram of the flow inside a precessing cylinder, for $h = 1.62$. For $\omega = 1.18$ (blue solid line at bottom) the first forced Kelvin mode is resonant, and the $(6,1,1)$ combination is unstable, see [39]. Extra curves correspond to the prediction (38) and show the stability threshold for non-resonant frequencies: $\omega = 1.25$ (red dashed line, combination $(6,1,1)$), $\omega = 1.3$ (yellow dotted line, combination $(5,1,1)$), $\omega = 1.35$ (purple dash-dotted line, combination $(5,1,1)$) and $\omega = 1.4$ (green solid line at top, combination $(5,1,1)$). Symbols are from Fig. 10 of [39] and represent the predictions of the resonant forcing theory for $Ek = 1.5 \times 10^{-4}$.

The critical Rossby number for which the instability appears can be determined from the condition of a vanishing real part of $\sigma$, leading to

$$|Ro_{\text{crit}}| = \left\{ \frac{\alpha_1^{\text{r}} \alpha_2^{\text{r}}}{c_1 c_2} \left[ 1 + \left( \frac{\alpha_1^{\text{i}} - \alpha_2^{\text{i}}}{\alpha_1^{\text{r}} + \alpha_2^{\text{r}}} \right)^2 \right] \right\}^{1/2}, \tag{38}$$

where $\alpha_l^{\text{r}}$ and $\alpha_l^{\text{i}}$ are respectively the real and imaginary parts of $\alpha_l$. For large $Ek$, the volume viscous effects (which scale as $Ek$) are larger than the surface viscous effects (which scale as $Ek^{1/2}$), such that $Ro_{\text{crit}}$ scales as $Ek^{3/2}$. On the contrary, for low $Ek$, surface viscous effects are dominant and $Ro_{\text{crit}}$ scales as $Ek$ [39].

Figure 4 shows the critical Rossby number (Eq.(38)) as a function of the Ekman number, for a cylinder with an aspect ratio $h = 1.62$. For $\omega = 1.18$, the first forced Kelvin mode is resonant: the amplitude of the base flow is maximum and the combination $(6,1,1)$ (which is exactly resonant) is the most unstable, see [39]. Outside of the resonance, the amplitude of the base flow is less intense and the free Kelvin modes are not exactly resonant. It follows that a higher Rossby number is required to reach an unstable state. We note that the critical Rossby number does not necessarily increase as $\omega$ is moved further away from the resonance. This is actually due to the fact that the terms entering in the definition of the instability coefficients $c_l$ change their sign as $\omega$ is varied. In other words, depending on $\omega$ these terms either damp the instability or pump it up. Finally, Fig. 4 shows that the off-resonance prediction (38) agrees with the resonant forcing theory of [39] as $\omega$ remains close to the resonant frequency.

In what follows we investigate the behavior of the system for a precessing frequency far away from a resonance.

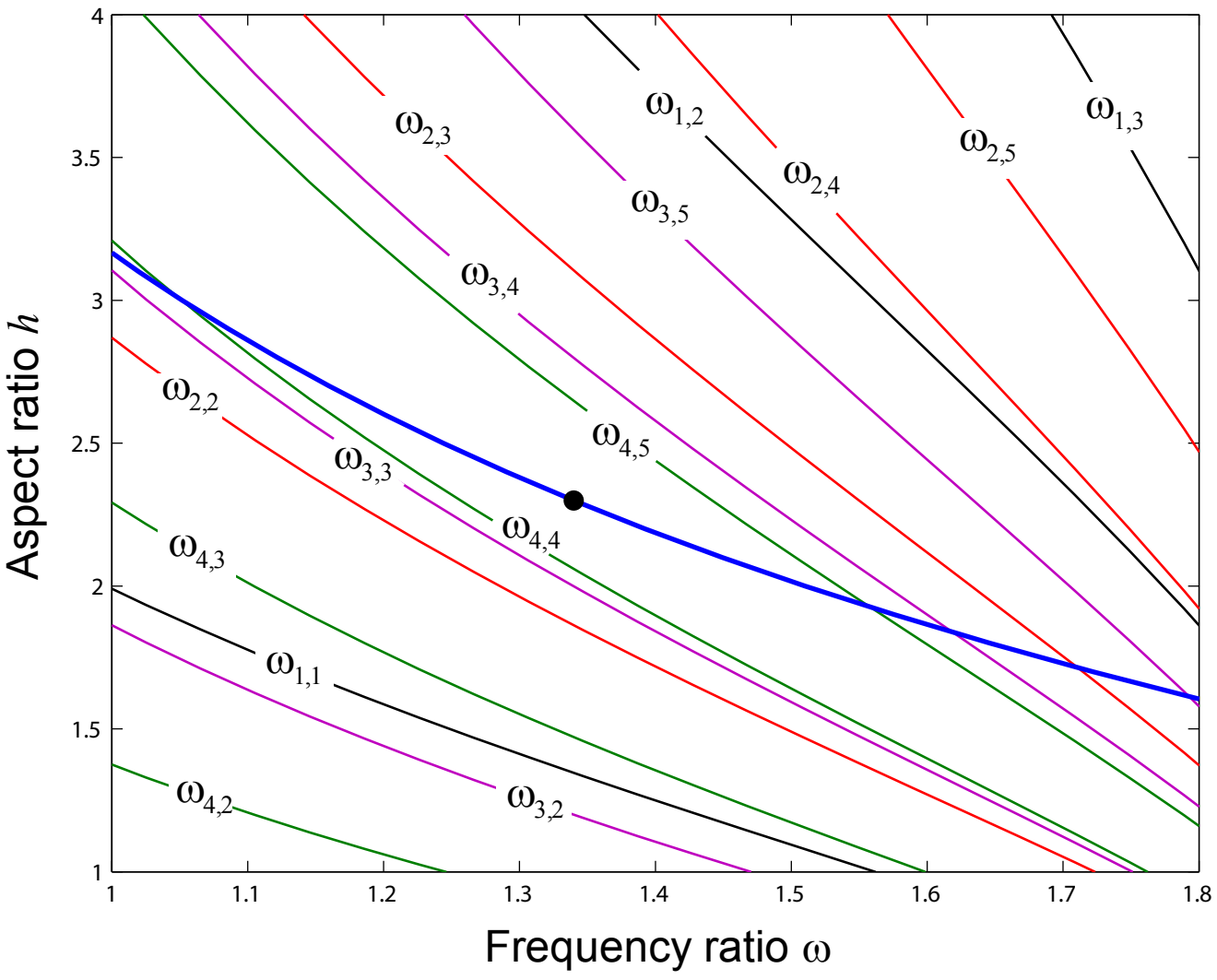

Figure 5. Resonance location in the $\omega$–$h$ plane. The resonances of the 1st (black curves), 2nd (red curves), 3rd (pink curves), 4th (green curves) forced modes are shown. The notation $\omega_{j,n}$ refers to the $n$-th resonance of the $j$-th forced mode. The blue thick line indicates the values of $\omega$ and $h$ for which the unstable modes combination $(3, 1, 1)$ forms a perfect triadic resonance with the first forced mode (see Fig. 3). The black circle ($\omega = 1.34, h = 2.3$) is the choice for the numerical application.

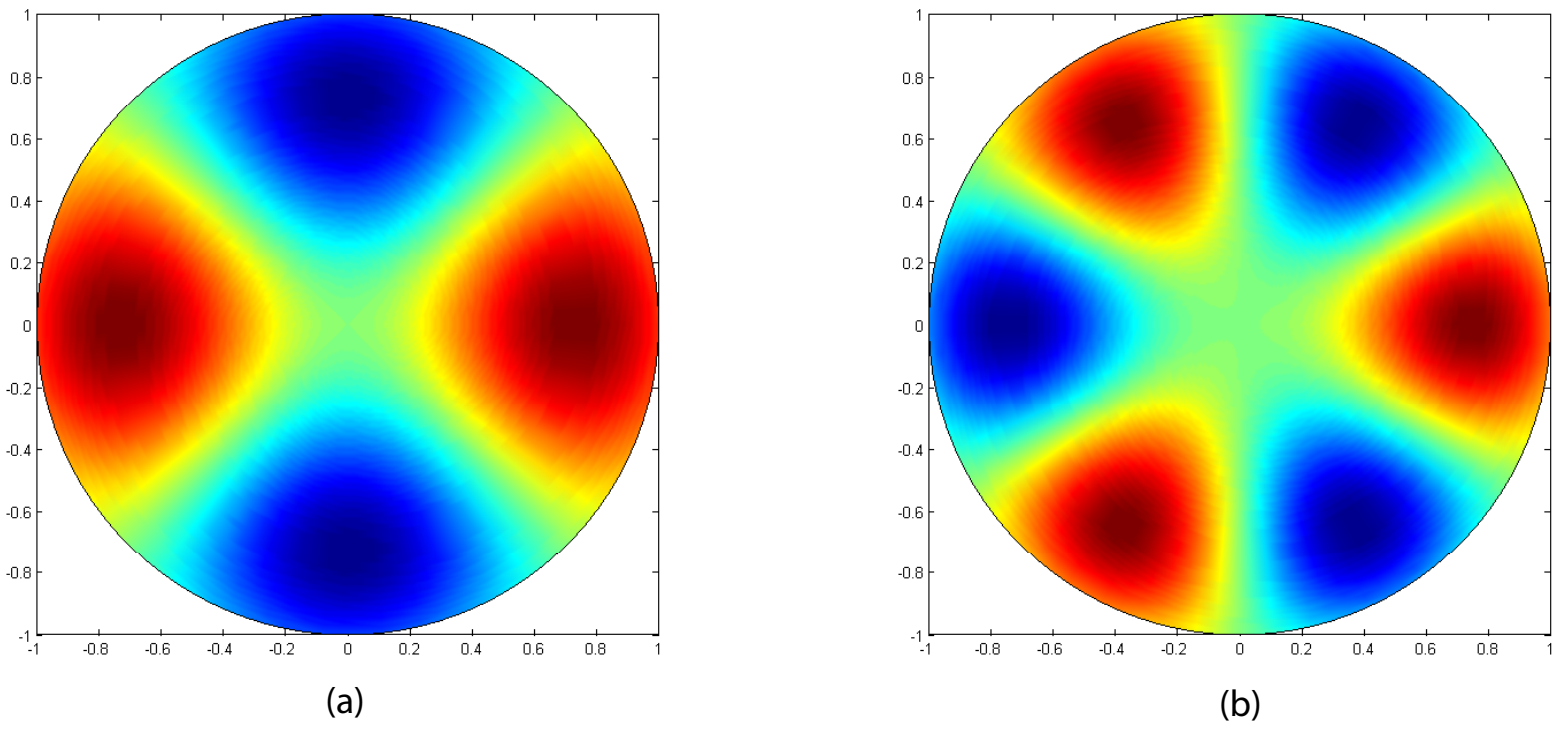

Figure 6. Theoretical axial vorticity fields of the free Kelvin modes $m_1 = 2$ (a) and $m_2 = 3$ (b) which are expected to grow for $h = 2.3$ and $\omega = 1.34$.

## 4. Numerical application far from a resonance

We first determine an aspect ratio $h$ and a frequency ratio $\omega$, such that the first 4 forced modes are away from a resonance, and such that combination of free Kelvin modes ($m_2 = 3, l_1 = 1, l_2 = 1$) forms a perfect triadic resonance with the first forced mode (see Fig. 3). These two conditions are illustrated in Fig. 5: the first 5 resonances of the first 4 forced modes are noted $\omega_{j,n}$ ($n$-th resonance of the $j$-th forced Kelvin mode) and are solutions to the dispersion relation

$$D\left(m_j, \omega_{j,n}, \frac{(4 - {\omega_{j,n}}^2)^{1/2}}{|\omega_{j,n}|}(2n-1)\frac{\pi}{h}\right) = 0. \tag{39}$$

The thick blue curve in Fig. 5 gives the aspect ratio and the frequency ratio for which the combination $(3, 1, 1)$ is resonant. This curve is the solution of the equation

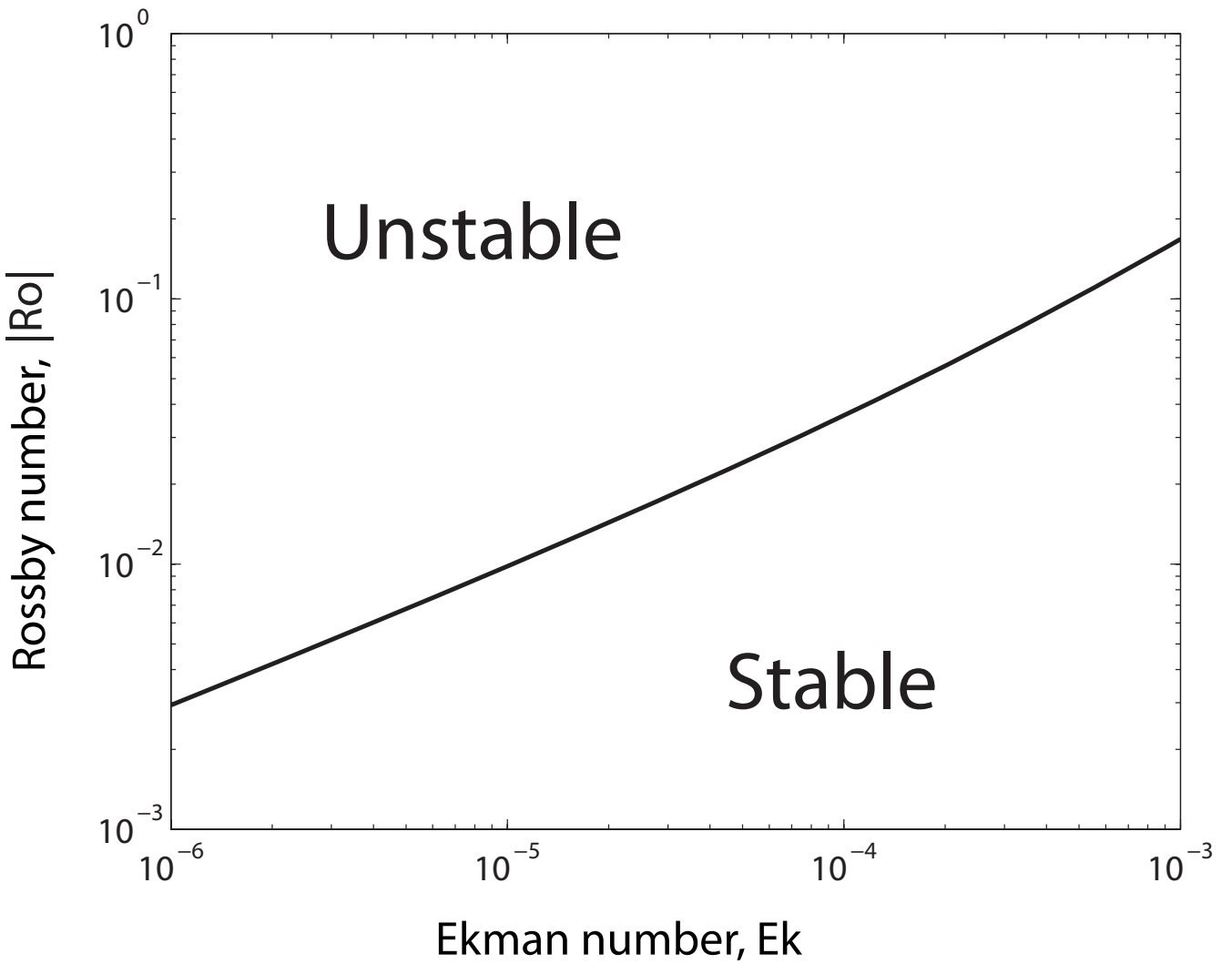

Figure 7. Stability diagram of the flow inside a precessing cylinder, for $h = 2.30$ and $\omega = 1.34$. The stable and unstable domains are separated by the solid line corresponding to the prediction (38).

| $\langle \mathbf{v}_1, \mathcal{I}\mathbf{v}_1 \rangle$ | $\bar{d}_{12}$ | $n_{1s}$ | $n_{1j1}$ | $n_{1j2}$ | $c_1$ | $s_1$ | $v_1$ |
|---|---|---|---|---|---|---|---|
| 27.1230 | 6.8448 | 1.1623 | −12.7524 | 3.6628 | −0.3940 | 1.22 − 0.15i | 34.37 |
| $\langle \mathbf{v}_2, \mathcal{I}\mathbf{v}_2 \rangle$ | $d_{21}$ | $n_{2s}$ | $n_{2j1}$ | $n_{2j2}$ | $c_2$ | $s_2$ | $v_2$ |
| 35.2671 | −6.8448 | 21.8624 | −23.9177 | 6.8697 | −0.5683 | 1.50 + 0.027i | 39.08 |

Table 1
Values of the parameters appearing in the amplitude equations. The aspect ratio and the frequency ratio are $h = 2.30$ and $\omega = 1.34$. For these values, the combination $(m_2 = 3, l_1 = 1, l_2 = 1)$ is resonant and corresponds to a pair of free Kelvin modes $(m_1 = 2, \omega_1 = -0.466, k_1 = \pi/h)$ and $(m_2 = 3, \omega_2 = 0.874, k_2 = 2\pi/h)$. The forced modes are those with indices $j_1 = 1$, $j_2 = 2$ and have axial wavenumbers $k_{j_1} = \pi/h$ and $k_{j_2} = 3\pi/h$. Their amplitudes are $a^{\mathrm{H}}_{j_1} = 1.4983$ and $a^{\mathrm{H}}_{j_2} = 0.1130$.

$$\omega_2(k_2 = 2\pi/h, l_2 = 1, m_2 = 3) = \omega_1(k_1 = \pi/h, l_1 = 1, m_{1r} = 2) + \omega_{i,n}, \tag{40}$$

where $\omega_2(k_2 = 2\pi/h, l_2 = 1, m_2 = 3)$ means: the value of $\omega_2$ for $k_2 = 2\pi/h$ and $l_2 = 1$ (first branch of the dispersion relation $m_2 = 3$).

In the present numerical investigations, we choose $h = 2.3$ and $\omega = 1.34$. It can be checked that for these aspect and frequency ratios, $(3, 1, 1)$ is a resonant point (Fig. 3). The theoretical axial vorticity fields of the corresponding free Kelvin modes, $m_1 = 2$ and $m_2 = 3$, are shown in Fig. 6. Since these modes correspond to the first branches of the dispersion relation ($l_1 = l_2 = 1$), their vorticity fields exhibit only one ring of 4 (for $m_1 = 2$) or 6 (for $m_2 = 3$) counter-rotating vortices.

We have listed in Table 1 the values of the parameters needed to compute the instability growth rate and the critical Rossby number. It appears that the terms entering in the definition of the instability coefficients $c_l$ have different effects: the positive ones are stabilizing and the negative ones are destabilizing. From the values of $c_l$, it comes an inviscid growth rate $\sigma_r = |Ro|\left(\sqrt{c_1 c_2}\right) = 0.4732|Ro|$. Since the combination $(3, 1, 1)$ corresponds to free Kelvin modes with simple radial and axial structures, volume viscous damping (which scale as $k_l^2 + \delta_l^2$) is small, which make these modes the perfect candidates for an instability. The stability diagram of the resonant combination $(3, 1, 1)$ is shown in Fig. 7 in the *Ek–Ro* plane: in this graph, the prediction from (38) is represented by a solid line that splits the *Ek–Ro* plane into a stable and an unstable domain.

## 5. Conclusion

In this paper, the instability of a fluid inside a precessing cylinder has been addressed theoretically. First, we have shown that the base flow can be written as a superposition of a vertical or horizontal shear flow and a sum of forced modes. The stability of this base flow has then been studied for a forcing at a non-resonant frequency, thus completing previous studies performed for resonant or near-resonant frequencies [37, 39]. We have shown that the non-resonant base flow can trigger a triadic instability with two free Kelvin modes only if these modes have different axial parities. We also have obtained a prediction of the instability growth rate, showing that the inviscid growth rate is proportional to the Rossby number, $Ro$, and an order $Ek^{-1/2}$ smaller than the growth rate obtained for a resonant base flow. Finally, introducing viscous damping, we have given a prediction of $Ro_{\text{crit}}$, the critical Rossby number for which the flow becomes unstable, as a function of $Ek$: for large (resp. low) $Ek$ numbers, $Ro_{\text{crit}}$ scales as $Ek^{3/2}$ (resp. $Ek$). Outside of the resonance, the amplitude of the base flow is less intensive and higher Rossby numbers are required for an instability.

We believe that the theoretical predictions provided in this paper will be valuable for future experimental and numerical studies of precessing flows performed at arbitrary frequencies. We shall note however that our analysis relies on the assumption of a small Rossby number, i.e. a small precession angle. For strong forcing, very different phenomena are expected (e.g. Kelvin-Helmholtz, centrifugal, or boundary layer instabilities), due to the generation of powerful zonal flows.

In closing, the precessional instability is typical of transition to turbulence in rotating flows. The presence of rotation ensures that energy is continuously provided to the flow. It also supports the existence of inertial waves that can lead to several instabilities (elliptic instability, libration instability, etc.). The structure of turbulence is also modified by the presence of the rotation because of the anisotropy induced. There is much more work to be done on this fascinating topic if we want to understand the mechanisms at play in turbulent rotating flows.

This study was carried out under CEA-CNRS contract No. 012171.

## Acknowledgment

The authors would like to thank the anonymous reviewer for his/her valuable comments and suggestions to improve the manuscript.

## Appendix A. Operators of the Euler equations, forced mode vector $\mathbf{v}_j$, amplitudes $a_j^{\mathbf{V}}$ and $a_j^{\mathbf{H}}$, free Kelvin modes vectors $\mathbf{v}_l^+$ and $\mathbf{v}_l^-$

Operators used for the matrix formulation (3) of the Euler equations are

$$\mathcal{I} = \begin{pmatrix} 1 & 0 & 0 & 0 \\ 0 & 1 & 0 & 0 \\ 0 & 0 & 1 & 0 \\ 0 & 0 & 0 & 0 \end{pmatrix}, \mathcal{D} = \begin{pmatrix} 0 & 0 & -\mathrm{i} & 0 \\ 0 & 0 & 1 & 0 \\ \mathrm{i} & -1 & 0 & 0 \\ 0 & 0 & 0 & 0 \end{pmatrix}, \mathcal{M} = \begin{pmatrix} 0 & -2 & 0 & \dfrac{\partial}{\partial r} \\ 2 & 0 & 0 & \dfrac{1}{r}\dfrac{\partial}{\partial \varphi} \\ 0 & 0 & 0 & \dfrac{\partial}{\partial z} \\ \dfrac{\partial}{\partial r} + \dfrac{1}{r} & \dfrac{1}{r}\dfrac{\partial}{\partial \varphi} & \dfrac{\partial}{\partial z} & 0 \end{pmatrix}, \tag{A.1}$$

and

$$\mathbf{F}_0 = \begin{pmatrix} 0 \\ 0 \\ -r\omega \\ 0 \end{pmatrix}, \mathbf{N}\left(\mathbf{v}_1, \mathbf{v}_2\right) = \begin{pmatrix} \mathbf{u}_1 \times \left(\nabla \times \mathbf{u}_2\right) \\ 0 \end{pmatrix}. \tag{A.2}$$

The forced modes quadri-vector $\mathbf{v}_j^{\mathrm{V}}$ and $\mathbf{v}_j^{\mathrm{H}}$ are given by the same expression

$$\mathbf{v}_j^{\mathrm{V}} = \mathbf{v}_j^{\mathrm{H}} = 2 \begin{pmatrix} \left(\mathbf{w}_{1,\omega,k_j}\right)_r \mathrm{i} \sin\left(k_j z\right) \\ \left(\mathbf{w}_{1,\omega,k_j}\right)_\varphi \mathrm{i} \sin\left(k_j z\right) \\ \left(\mathbf{w}_{1,\omega,k_j}\right)_z \cos\left(k_j z\right) \\ \left(\mathbf{w}_{1,\omega,k_j}\right)_p \mathrm{i} \sin\left(k_j z\right) \end{pmatrix}, \tag{A.3}$$

with

$$\mathbf{w}_{m,\omega,k} = \begin{pmatrix} \dfrac{-1}{4-\omega^2}\left(\omega\delta J_m^{'}\left(\delta r\right) + 2\dfrac{m}{r} J_m\left(\delta r\right)\right) \\ \dfrac{-\mathrm{i}}{4-\omega^2}\left(2\delta J_m^{'}\left(\delta r\right) + \dfrac{\omega m}{r} J_m\left(\delta r\right)\right) \\ \mathrm{i}\dfrac{k}{\omega} J_m\left(\delta r\right) \\ -\mathrm{i} J_m\left(\delta r\right) \end{pmatrix}, \tag{A.4}$$

and where $J_m$ is the Bessel function of the first kind, $J_m^{'}$ its derivative and $\delta = \sqrt{4-\omega^2}k/|\omega|$. However, they have been named differently since they have different boundary conditions and thus different axial wavenumbers $k_j^{\mathrm{V}}$ or $k_j^{\mathrm{H}}$. For the vertical shear, the radial wavenumber $\delta$ is imposed by the boundary conditions and $k$ is deduced from the dispersion relation whereas for the horizontal shear, the axial wavenumber is imposed by the boundary conditions and $\delta$ is given again by $\delta = \sqrt{4-\omega^2}k/|\omega|$.

The dispersion relation is

$$D(m,\omega,k) = \omega\delta J_m^{'}\left(\delta\right) + 2mJ_m\left(\delta\right). \tag{A.5}$$

The amplitudes of the forced modes are

$$a_j^{\mathrm{V}} = \frac{\omega^2}{(\omega-2)(k_j^2+1)k_j J_1(\delta_j)\cos(k_j h/2)}, \tag{A.6a}$$

$$a_j^{\mathrm{H}} = -\frac{2h(-1)^j\omega(2+\omega)}{\pi^2(2j-1)^2(\omega\delta_j J_1'(\delta_j) + 2J_1(\delta_j))}. \tag{A.6b}$$

Vectors $\mathbf{v}_l^+$ and $\mathbf{v}_l^-$ appearing in (20) are

$$\mathbf{v}_l^+ = 2 \begin{pmatrix} \left(\mathbf{w}_{m_l,\omega_l,k_l^+}\right)_r \cos\left(k_l^+ z\right) \\ \left(\mathbf{w}_{m_l,\omega_l,k_l^+}\right)_\varphi \cos\left(k_l^+ z\right) \\ \left(\mathbf{w}_{m_l,\omega_l,k_l^+}\right)_z \mathrm{i}\sin\left(k_l^+ z\right) \\ \left(\mathbf{w}_{m_l,\omega_l,k_l^+}\right)_p \cos\left(k_l^+ z\right) \end{pmatrix}, \quad \mathbf{v}_l^- = 2 \begin{pmatrix} \left(\mathbf{w}_{m_l,\omega_l,k_l^-}\right)_r \mathrm{i}\sin\left(k_l^- z\right) \\ \left(\mathbf{w}_{m_l,\omega_l,k_l^-}\right)_\varphi \mathrm{i}\sin\left(k_l^- z\right) \\ \left(\mathbf{w}_{m_l,\omega_l,k_l^-}\right)_z \cos\left(k_l^- z\right) \\ \left(\mathbf{w}_{m_l,\omega_l,k_l^-}\right)_p \mathrm{i}\sin\left(k_l^- z\right) \end{pmatrix}. \tag{A.7}$$

## Appendix B. Derivation of amplitude equations

In this Appendix, we derive the amplitude equations (26), starting from the order *Ro* equation (25) that we report here

$$\left(\frac{\partial \mathcal{I}}{\partial t} + \mathcal{M}\right) \widetilde{\mathbf{v}}_1 = \mathbf{N}\left(\mathbf{v}_{\text{base}}, \widetilde{\mathbf{v}}_0\right) + \mathbf{N}\left(\widetilde{\mathbf{v}}_0, \mathbf{v}_{\text{base}}\right) + \left[\left(\mathcal{D}e^{\mathrm{i}(\omega t+\varphi)} + \text{c.c.}\right) - \frac{\partial \mathcal{I}}{\partial \tau}\right] \widetilde{\mathbf{v}}_0. \tag{B.1}$$

As explained in the core of the manuscript, a solvability condition is obtained by taking the dot product of this equation with $\mathbf{v}_l e^{\mathrm{i}(\omega_l t + m_j\varphi)}$, $l = 1, 2$. The problem being self adjoint, we have $\left\langle \mathbf{v}_l e^{\mathrm{i}(\omega_l t+m_l\varphi)}, \left(\partial\mathcal{I}/\partial t + \mathcal{M}\right) \widetilde{\mathbf{v}}_1 \right\rangle = 0$, so that, we are left with

$$\begin{aligned}\left\langle \mathbf{v}_l e^{\mathrm{i}(\omega_l t+m_l\varphi)}, \frac{\partial \mathcal{I}}{\partial \tau}\widetilde{\mathbf{v}}_0 \right\rangle = \left\langle \mathbf{v}_l e^{\mathrm{i}(\omega_l t+m_l\varphi)}, \left(\mathcal{D}e^{\mathrm{i}(\omega t+\varphi)} + \text{c.c.}\right)\widetilde{\mathbf{v}}_0 \right\rangle + \left\langle \mathbf{v}_l e^{\mathrm{i}(\omega_l t+m_l\varphi)}, \mathbf{N}\left(\mathbf{v}_{\text{base}}, \widetilde{\mathbf{v}}_0\right) \right\rangle \\ + \left\langle \mathbf{v}_l e^{\mathrm{i}(\omega_l t+m_l\varphi)}, \mathbf{N}\left(\widetilde{\mathbf{v}}_0, \mathbf{v}_{\text{base}}\right) \right\rangle.\end{aligned} \tag{B.2a}$$

The computation of the LHS term is straightforward and follows from the linearity of the dot product and the orthogonality of the free Kelvin modes. Introducing the expression of $\widetilde{\mathbf{v}}_0$ given by (21) gives a LHS term equal to

$$\left\langle \mathbf{v}_l e^{\mathrm{i}(\omega_l t+m_l\varphi)}, \sum_{l=1}^{2} \frac{dA_j}{d\tau}\mathcal{I}\left(\mathbf{v}_j e^{\mathrm{i}(\omega_j t+m_j\varphi)}\right) \right\rangle + \left\langle \mathbf{v}_l e^{\mathrm{i}(\omega_l t+m_l\varphi)}, \sum_{j=1}^{2} \frac{d\overline{A_j}}{d\tau}\mathcal{I}\left(\overline{\mathbf{v}_j} e^{-\mathrm{i}(\omega_j t+m_j\varphi)}\right) \right\rangle, \tag{B.3a}$$

$$= \sum_{j=1}^{2} \frac{dA_j}{d\tau} \left\langle \mathbf{v}_l e^{\mathrm{i}(\omega_l t+m_l\varphi)}, e^{\mathrm{i}(\omega_j t+m_j\varphi)}\mathcal{I}\mathbf{v}_j \right\rangle + \sum_{j=1}^{2} \frac{d\overline{A_j}}{d\tau} \left\langle \mathbf{v}_l e^{\mathrm{i}(\omega_l t+m_l\varphi)}, e^{-\mathrm{i}(\omega_j t+m_j\varphi)}\mathcal{I}\overline{\mathbf{v}_j} \right\rangle, \tag{B.3b}$$

$$= \sum_{j=1}^{2} \frac{dA_j}{d\tau} \underbrace{\left\langle \mathbf{v}_l, e^{\mathrm{i}[(-\omega_l+\omega_j)t+(-m_l+m_j)\varphi]}\mathcal{I}\mathbf{v}_j \right\rangle}_{\neq 0 \text{ if } j=l} + \sum_{j=1}^{2} \frac{d\overline{A_j}}{d\tau} \underbrace{\left\langle \mathbf{v}_l, e^{\mathrm{i}[(-\omega_l-\omega_j)t+(-m_l-m_j)\varphi]}\mathcal{I}\overline{\mathbf{v}_j} \right\rangle}_{0 \text{ because } \propto \int_0^{2\pi} e^{-\mathrm{i}(m_l+m_j)\varphi}d\varphi}, \tag{B.3c}$$

$$= \frac{dA_l}{d\tau} \left\langle \mathbf{v}_l, \mathcal{I}\mathbf{v}_l \right\rangle. \tag{B.3d}$$

Eq. (B.3c) shows that it is not necessary to take into account the c.c. part of $A_j \mathbf{v}_j e^{\mathrm{i}(\omega_j t+m_j\varphi)}$ in (B.2) since it leads to 0 integral terms. This observation still holds for computations with operators $\mathcal{D}$, $\overline{\mathcal{D}}$ and $\mathbf{N}$ since they do not change the wavenumbers in the exponential when applied to $\mathbf{v}_l e^{\mathrm{i}(\omega_l t+m_l\varphi)}$. Therefore, the c.c. part of $A_j \mathbf{v}_j e^{\mathrm{i}(\omega_j t+m_j\varphi)}$ will be omitted in the next computations.

Plugging (B.3d) into (B.2) yields the amplitude equations

$$\frac{dA_l}{d\tau} = \frac{\left\langle \mathbf{v}_l e^{\mathrm{i}(\omega_l t+m_l\varphi)}, \left(\mathcal{D}e^{\mathrm{i}(\omega t+\varphi)} + \text{c.c.}\right)\widetilde{\mathbf{v}}_0 \right\rangle + \left\langle \mathbf{v}_l e^{\mathrm{i}(\omega_l t+m_l\varphi)}, \mathbf{N}\left(\mathbf{v}_{\text{base}}, \widetilde{\mathbf{v}}_0\right) + \mathbf{N}\left(\widetilde{\mathbf{v}}_0, \mathbf{v}_{\text{base}}\right) \right\rangle}{\left\langle \mathbf{v}_l, \mathcal{I}\mathbf{v}_l \right\rangle}. \tag{B.4}$$

We now proceed with the calculation of the RHS terms of (B.4). Computations are performed with $l = 1$ so that results for $l = 2$ will follow from the permutation of indices $(1, 2) \to (2, 1)$. Also, from the operators properties presented in §2.2, we know that the free Kelvin modes $\mathbf{v}_1 e^{\mathrm{i}(\omega_1 t+m_1\varphi)}$ and $\mathbf{v}_2 e^{\mathrm{i}(\omega_2 t+m_2\varphi)}$ must have different $z$-parities to give nonzero coupling terms. Thus, for $l = 1$ we can directly substitute in (B.4) vector $\widetilde{\mathbf{v}}_0$ by $\mathbf{v}_2 e^{\mathrm{i}(\omega_2 t+m_2\varphi)}$ and (B.4) writes

$$\frac{dA_1}{d\tau} = \frac{\left\langle \mathbf{v}_1 e^{\mathrm{i}(\omega_1 t + m_1\varphi)}, \left(\mathcal{D} e^{\mathrm{i}(\omega t+\varphi)} + \text{c.c.}\right) \mathbf{v}_2 e^{\mathrm{i}(\omega_2 t + m_2\varphi)} \right\rangle}{\langle \mathbf{v}_1, \mathcal{I}\mathbf{v}_1 \rangle} + \frac{\left\langle \mathbf{v}_1 e^{\mathrm{i}(\omega_1 t + m_1\varphi)}, \mathbf{N}\left(\mathbf{v}_{\text{base}}, \mathbf{v}_2 e^{\mathrm{i}(\omega_2 t + m_2\varphi)}\right)\right\rangle}{\langle \mathbf{v}_1, \mathcal{I}\mathbf{v}_1 \rangle}$$
$$+ \frac{\left\langle \mathbf{v}_1 e^{\mathrm{i}(\omega_1 t + m_1\varphi)}, \mathbf{N}\left(\mathbf{v}_2 e^{\mathrm{i}(\omega_2 t + m_2\varphi)}, \mathbf{v}_{\text{base}}\right)\right\rangle}{\langle \mathbf{v}_1, \mathcal{I}\mathbf{v}_1 \rangle}. \tag{B.5a}$$

In what follows we compute each of the terms in the RHS of (B.5a) and we assume that the resonance conditions (24) are fulfilled in order to drop the exponential terms.

B.1. *Computation of* $\left\langle \mathbf{v}_1 e^{i(\omega_1 t + m_1\varphi)}, \left(\mathcal{D} e^{i(\omega t+\varphi)} + \text{c.c.}\right) \left(\mathbf{v}_2 e^{i(\omega_2 t + m_2\varphi)}\right)\right\rangle$

Expanding the complex conjugate leads to 2 terms:

$$\left\langle \mathbf{v}_1 e^{\mathrm{i}(\omega_1 t + m_1\varphi)}, e^{\mathrm{i}(\omega t+\varphi)} e^{\mathrm{i}(\omega_2 t + m_2\varphi)} \mathcal{D}\mathbf{v}_2 \right\rangle + \left\langle \mathbf{v}_1 e^{\mathrm{i}(\omega_1 t + m_1\varphi)}, e^{-\mathrm{i}(\omega t+\varphi)} e^{\mathrm{i}(\omega_2 t + m_2\varphi)} \overline{\mathcal{D}}\mathbf{v}_2 \right\rangle,$$

The first term vanishes because the azimuthal Fourier components are different on each side of the dot product such that the integral over $\varphi$ gives zero. In contrast, in the second term, the azimuthal Fourier components are equal and can thus be dropped. This term can thus be written as

$$\overline{d}_{12} = \langle \mathbf{v}_1, \overline{\mathcal{D}}\mathbf{v}_2 \rangle. \tag{B.6}$$

B.2. *Computation of* $\left\langle \mathbf{v}_1 e^{i(\omega_1 t + m_1\varphi)}, \mathbf{N}\left(\mathbf{v}_{\text{base}}, \mathbf{v}_2 e^{i(\omega_2 t + m_2\varphi)}\right)\right\rangle$

Here $\mathbf{v}_{\text{base}} = \mathbf{v}_b e^{\mathrm{i}(\omega t+\varphi)} + \text{c.c.}$ is the base flow given either by (8) or (9) depending on which decomposition (vertical or horizontal shear) is used to express the base flow. We have

$$\left\langle \mathbf{v}_1 e^{\mathrm{i}(\omega_1 t + m_1\varphi)}, \mathbf{N}\left(\mathbf{v}_{\text{base}}, \mathbf{v}_2 e^{\mathrm{i}(\omega_2 t + m_2\varphi)}\right)\right\rangle \tag{B.7a}$$

$$= \left\langle \mathbf{v}_1 e^{\mathrm{i}(\omega_1 t + m_1\varphi)}, \mathbf{N}\left(\mathbf{v}_b e^{\mathrm{i}(\omega t+\varphi)} + \text{c.c.}, \mathbf{v}_2 e^{\mathrm{i}(\omega_2 t + m_2\varphi)}\right)\right\rangle, \tag{B.7b}$$

$$= \left\langle \mathbf{v}_1 e^{\mathrm{i}(\omega_1 t + m_1\varphi)}, \mathbf{N}\left(\mathbf{v}_b e^{\mathrm{i}(\omega t+\varphi)}, \mathbf{v}_2 e^{\mathrm{i}(\omega_2 t + m_2\varphi)}\right)\right\rangle + \left\langle \mathbf{v}_1 e^{\mathrm{i}(\omega_1 t + m_1\varphi)}, \mathbf{N}\left(\overline{\mathbf{v}_b} e^{-\mathrm{i}(\omega t+\varphi)}, \mathbf{v}_2 e^{\mathrm{i}(\omega_2 t + m_2\varphi)}\right)\right\rangle, \tag{B.7c}$$

$$= \left\langle \mathbf{v}_1 e^{\mathrm{i}(\omega_1 t + m_1\varphi)}, e^{\mathrm{i}(\omega t+\varphi)} e^{\mathrm{i}(\omega_2 t + m_2\varphi)} \mathbf{N}_{\mathrm{i}m_2}\left(\mathbf{v}_b, \mathbf{v}_2\right)\right\rangle$$

$$+ \left\langle \mathbf{v}_1 e^{\mathrm{i}(\omega_1 t + m_1\varphi)}, e^{-\mathrm{i}(\omega t+\varphi)} e^{\mathrm{i}(\omega_2 t + m_2\varphi)} \mathbf{N}_{\mathrm{i}m_2}\left(\overline{\mathbf{v}_b}, \mathbf{v}_2\right)\right\rangle, \tag{B.7d}$$

(B.7e)

where $\mathbf{N}_{\mathrm{i}m_2}$ corresponds to operator $\mathbf{N}$ where $d/d\varphi$ has been replaced by $\mathrm{i}m_2$. As previously, the first term vanishes and the exponential can be dropped from the second term. Introducing the expression of $\mathbf{v}_b$ makes this whole term equal to

$$\left\langle \mathbf{v}_1, \mathbf{N}_{\mathrm{i}m_2}\left(\overline{\mathbf{v}_s} + \sum_{j=1}^{\infty} \overline{a_j}\,\overline{\mathbf{v}_j}, \mathbf{v}_2\right)\right\rangle = n_{1\bar{s}2} + \sum_{j=1}^{\infty} \overline{a_j} n_{1\bar{j}2}, \tag{B.8a}$$

$$\text{with} \quad n_{1\bar{s}2} = \left\langle \mathbf{v}_1, \mathbf{N}_{\mathrm{i}m_2}\left(\overline{\mathbf{v}_s}, \mathbf{v}_2\right)\right\rangle \quad \text{and} \quad n_{1\bar{j}2} = \left\langle \mathbf{v}_1, \mathbf{N}_{\mathrm{i}m_2}\left(\overline{\mathbf{v}_j}, \mathbf{v}_2\right)\right\rangle, \tag{B.8b}$$

B.3. *Computation of* $\left\langle \mathbf{v}_1 e^{i(\omega_1 t + m_1\varphi)}, \mathbf{N}\left(\mathbf{v}_2 e^{i(\omega_2 t + m_2\varphi)}, \mathbf{v}_{\text{base.}}\right)\right\rangle$

We have

$$\left\langle \mathbf{v}_1 e^{\mathrm{i}(\omega_1 t + m_1\varphi)}, \mathbf{N}\left(\mathbf{v}_2 e^{\mathrm{i}(\omega_2 t + m_2\varphi)}, \mathbf{v}_{\text{base}}\right)\right\rangle \tag{B.9a}$$

$$= \left\langle \mathbf{v}_1 e^{\mathrm{i}(\omega_1 t + m_1\varphi)}, \mathbf{N}\left(\mathbf{v}_2 e^{\mathrm{i}(\omega_2 t + m_2\varphi)}, \mathbf{v}_b e^{\mathrm{i}(\omega t + \varphi)} + \text{c.c.}\right)\right\rangle, \tag{B.9b}$$

$$= \left\langle \mathbf{v}_1 e^{\mathrm{i}(\omega_1 t + m_1\varphi)}, \mathbf{N}\left(\mathbf{v}_2 e^{\mathrm{i}(\omega_2 t + m_2\varphi)}, \mathbf{v}_b e^{\mathrm{i}(\omega t + \varphi)}\right)\right\rangle + \left\langle \mathbf{v}_1 e^{\mathrm{i}(\omega_1 t + m_1\varphi)}, \mathbf{N}\left(\mathbf{v}_2 e^{\mathrm{i}(\omega_2 t + m_2\varphi)}, \overline{\mathbf{v}_b} e^{-\mathrm{i}(\omega t + \varphi)}\right)\right\rangle, \tag{B.9c}$$

$$= \left\langle \mathbf{v}_1 e^{\mathrm{i}(\omega_1 t + m_1\varphi)}, e^{\mathrm{i}(\omega_2 t + m_2\varphi)} e^{\mathrm{i}(\omega t + \varphi)} \mathbf{N}_{\mathrm{i}}\left(\mathbf{v}_2, \mathbf{v}_b\right)\right\rangle + \left\langle \mathbf{v}_1 e^{\mathrm{i}(\omega_1 t + m_1\varphi)}, e^{\mathrm{i}(\omega_2 t + m_2\varphi)} e^{-\mathrm{i}(\omega t + \varphi)} \mathbf{N}_{-\mathrm{i}}\left(\mathbf{v}_2, \overline{\mathbf{v}_b}\right)\right\rangle, \tag{B.9d}$$

(B.9e)

where $\mathbf{N}_{\mathrm{i}}$ and $\mathbf{N}_{-\mathrm{i}}$ correspond to operator $\mathbf{N}$ where $d/d\varphi$ has been replaced by $\mathrm{i}$ and $-\mathrm{i}$, respectively. As previously, the first term vanishes and the exponentials can be dropped from the second term. Introducing the expression of $\mathbf{v}_b$ makes this whole term equal to

$$\left\langle \mathbf{v}_1, \mathbf{N}_{-\mathrm{i}}\left(\mathbf{v}_2, \overline{\mathbf{v}_s} + \sum_{j=1}^{\infty} \overline{a_j} \overline{\mathbf{v}_j}\right)\right\rangle = n_{12\overline{s}} + \sum_{j=1}^{\infty} \overline{a_j} n_{12\overline{j}}, \tag{B.10a}$$

$$\text{with} \quad n_{12\overline{s}} = \left\langle \mathbf{v}_1, \mathbf{N}_{-\mathrm{i}}\left(\mathbf{v}_2, \overline{\mathbf{v}_s}\right)\right\rangle \quad \text{and} \quad n_{12\overline{j}} = \left\langle \mathbf{v}_1, \mathbf{N}_{-\mathrm{i}}\left(\mathbf{v}_2, \overline{\mathbf{v}_j}\right)\right\rangle \tag{B.10b}$$

B.4. *Conditions of resonance and amplitude equations*

Collecting (B.6), (B.8b) and (B.10b) together, the amplitude equations (B.4) rewrite

$$\frac{dA_1}{d\tau} = A_2 \frac{\overline{d}_{12} + n_{1s} + \sum_{j=1}^{\infty} \overline{a_j} n_{1j}}{\langle \mathbf{v}_1, \mathcal{I}\mathbf{v}_1 \rangle}, \tag{B.11a}$$

$$\frac{dA_2}{d\tau} = A_1 \frac{d_{21} + n_{2s} + \sum_{j=1}^{\infty} a_j n_{2j}}{\langle \mathbf{v}_2, \mathcal{I}\mathbf{v}_2 \rangle}, \tag{B.11b}$$

where coefficients in (B.11a) are

$$\overline{d}_{12} = \langle \mathbf{v}_1, \overline{\mathcal{D}}\mathbf{v}_2 \rangle, \tag{B.12a}$$

$$n_{1s} = n_{1\overline{s}2} + n_{12\overline{s}} = \langle \mathbf{v}_1, \mathbf{N}_{\mathrm{i}m_2}(\overline{\mathbf{v}_s}, \mathbf{v}_2) \rangle + \langle \mathbf{v}_1, \mathbf{N}_{-\mathrm{i}}(\mathbf{v}_2, \overline{\mathbf{v}_s}) \rangle, \tag{B.12b}$$

$$n_{1j} = n_{1\overline{j}2} + n_{12\overline{j}} = \langle \mathbf{v}_1, \mathbf{N}_{\mathrm{i}m_2}(\overline{\mathbf{v}_j}, \mathbf{v}_2) \rangle + \langle \mathbf{v}_1, \mathbf{N}_{-\mathrm{i}}(\mathbf{v}_2, \overline{\mathbf{v}_j}) \rangle. \tag{B.12c}$$

Coefficients in (B.11b) are

$$d_{21} = \langle \mathbf{v}_2, \mathcal{D}\mathbf{v}_1 \rangle = -\overline{d}_{12}, \tag{B.13a}$$

$$n_{2s} = n_{2s1} + n_{21s} = \langle \mathbf{v}_2, \mathbf{N}_{\mathrm{i}m_1}(\mathbf{v}_s, \mathbf{v}_1) \rangle + \langle \mathbf{v}_2, \mathbf{N}_{\mathrm{i}}(\mathbf{v}_1, \mathbf{v}_s) \rangle, \tag{B.13b}$$

$$n_{2j} = n_{2j1} + n_{21j} = \langle \mathbf{v}_2, \mathbf{N}_{\mathrm{i}m_1}(\mathbf{v}_j, \mathbf{v}_1) \rangle + \langle \mathbf{v}_2, \mathbf{N}_{\mathrm{i}}(\mathbf{v}_1, \mathbf{v}_j) \rangle. \tag{B.13c}$$

## References

[1] K. Stewartson, On the stability of a spinning top containing liquid, J. Fluid Mech. 5 (1958) 577–592.

[2] B. G. Karpov, Ballistic Research Labs. Maryland, U. S., Report BRL R 1302 (1965).

[3] R. F. Gans, Dynamics of a near-resonant fluid-filled gyroscope, AIAA J. 22 (1984) 1465–1471.

[4] S. C. Garg, N. Furunoto, J. P. Vanyo, Spacecraft nutational instability prediction by energy dissipation measurments, J. Guid. 9 (1986) 357–361.

[5] T. Herbert, Viscous fluid motion in a spinning and nutating cylinder, J. Fluid Mech. 167 (1986) 181–198.

[6] J. J. Pocha, An experimental investigation of spacecraft sloshing, Space Commun. Broadcasting 5 (1987) 323–332.

[7] B. N. Agrawal, Dynamics characteristics of liquid motion in partially filled tanks of a spinning spacecraft, J. Guid. Control Dynam. 16 (1993) 636–640.

[8] G. W. Bao, M. Pascal, Stability of a spinning liquid filled spacecraft, Appl. Mech. 67 (1997) 407–421.

[9] J. P. Lambelin, F. Nadal, R. Lagrange, A. Sarthou, Non-resonant viscous theory for the stability of a fluid-filled gyroscope, Journal of Fluid Mechanics 639 (2009) 167–194.

[10] W. V. R. Malkus, An experimental study of global instabilities due to tidal (elliptical) distortion of a rotating elastic cylinder, Geophys. Astrophys. Fluid Dynamics 48 (1989) 123–134.

[11] F. Grote, H. Busse, A. Tilgner, Convection driven quadrupolar dynamos in rotating spherical shells, Phys. Rev. E 60 (1999) R5025.

[12] F. Grote, H. Busse, A. Tilgner, Regular and chaotic spherical dynamos, Phys. Earth Planet. Inter. 117 (2000) 259.

[13] F. Grote, H. Busse, Dynamics of convection and dynamos in rotating spherical fluid shells, Fluid Dyn. Res. 28 (2001) 349–368.

[14] D. Cébron, M. L. Bars, P. Maubert, Magnetohydrodynamic simulations of the elliptical instability in triaxial ellipsoids, Geophys. and Astrophys. Fluid Dyn. 106 (2011) 524–546.
[15] A. Tilgner, Precession driven dynamos, Phys. Fluids 17 (2005) 034104.
[16] C. Nore, J. Léorat, J. Guermond, F. Luddens, Nonlinear dynamo action in a precessing cylindrical container, Phys. Rev. E 84 (2011) 016317.
[17] M. G. Rochester, J. A. Jacobs, D. E. Smylie, K. F. Chong, Can precession power the geomagnetic dynamo?, Geophys. J. R. Astron. Soc. 43 (1975) 661.
[18] D. E. Loper, Torque balance and energy budget for the precesionally driven dynamo, Phys. Earth Planet. Inter. 43 (1975) 43.
[19] P. H. Roberts, D. Gubbins, Origin of the main field: Kinematics, Geomagnetism, Ed. J. A. Jacobs, Academic Press 2 (1987) 185–249.
[20] J. P. Vanyo, P. Wilde, P. Cardin, Experiments on precessing flows in the earth's liquid core, Geophys. J. Int. 121 (1995) 136–142.
[21] R. R. Kerswell, Upper bounds on the energy dissipation in turbulent precession,, J. Fluid Mech. 321 (1996) 335–370.
[22] R. F. Gans, On the precession of a resonant cylinder, J. Fluid Mech. 476 (1970) 865–872.
[23] A. Tilgner, Oscillatory shear layer in source driven flows in an unbounded rotating fluid, Phys. Fluids 12 (2000) 1101–1111.
[24] A. Gailitis, O. Lielaussis, E. Platacis, F. Stefani, G. Gerbeth, Laboratory experiments on hydromagnetic dynamos, Rev. Mod. Phys. 74 (2002) 973.
[25] L. Kelvin, Vibrations of a columnar vortex, Phil. Mag. 10 (1880) 155–168.
[26] H. P. Greenspan, The theory of rotating fluids, Cambridge University Press, 1968.
[27] D. Fultz, A note on overstability and elastoid-inertia oscillations of kelvin, solberg and bjerknes, J. Meteorol. 16 (1959) 199–208.
[28] A. D. McEwan, Inertial oscillations in a rotating fluid cylinder, J. Fluid Mech. 40 (1970) 603–640.
[29] R. R. Kerswell, C. F. Barenghi, On the viscous decay rates of inertial waves in a rotating cylinder, J. Fluid Mech. 285 (1995) 203–214.
[30] P. Meunier, C. Eloy, R. Lagrange, F. Nadal, A rotating fluid cylinder subject to weak precession, J. Fluid Mech. 599 (2008) 405–440.
[31] D. Kong, Z. Cui, X. Liao, K. Zhang, On the transition from the laminar to disordered flow in a precessing spherical-like cylinder, Geophys. Astrophys. Fluid Dyn. 109 (2015) 62–83.
[32] R. Thompson, Diurnal tides and shear instabilities in a rotating cylinder, J. Fluid Mech. 40 (1970) 737–751.
[33] R. Manasseh, Breakdown regimes of inertia waves in a precessing cylinder, J. Fluid Mech. 243 (1992) 261–296.
[34] R. Manasseh, Distorsions of inertia waves in a precessing cylinder forced near its fundamental mode resonance, J. Fluid Mech. 265 (1994) 345–370.
[35] J. J. Kobine, Inertial wave dynamics in a rotating and precessing cylinder, J. Fluid Mech. 303 (1995) 233–252.
[36] R. Manasseh, Nonlinear behaviour of contained inertia waves, J. Fluid Mech. 315 (1996) 151–173.
[37] R. Lagrange, C. Eloy, F. Nadal, P. Meunier, Instability of a fluid inside a precessing cylinder, Physics of Fluids. 20(8) (2008) 081701.
[38] R. Lagrange, P. Meunier, C. Eloy, F. Nadal, Dynamics of a fluid inside a precessing cylinder, Mechanics and Industry. 10 (2009) 187–194.
[39] R. Lagrange, P. Meunier, F. Nadal, C. Eloy, Precessional instability of a fluid cylinder, Journal of Fluid Mechanics 666 (2011) 104–145.